\documentclass[aps, prx,notitlepage,reprint,amsmath,amssymb,superscriptaddress, twocolumn, 10pt]{revtex4-2}

\pdfoutput=1

\usepackage{dcolumn}
\usepackage{bm}
\usepackage{amsmath}
\usepackage{amssymb,amsthm}
\usepackage{bbm}
\usepackage{epsfig,color}
\usepackage{scalerel}
\usepackage[sans]{dsfont}
\usepackage{comment}
\PassOptionsToPackage{hyphens}{url}
\usepackage{hyperref}
\usepackage{url}
\usepackage{units}
\usepackage{color}
\usepackage{ifthen}
\usepackage{graphicx}
\graphicspath{ {images/} }

\usepackage{enumerate}

\usepackage{mathrsfs}

\usepackage{tikz-cd}
\usepackage[utf8]{inputenc}
\usepackage{tikz}
\usetikzlibrary{automata,positioning}
\usetikzlibrary{fit,shapes.geometric}
\usetikzlibrary{decorations.pathmorphing}
\usetikzlibrary{arrows,matrix}
\usepackage{tikz-3dplot}
\usetikzlibrary{patterns}
\usetikzlibrary{arrows.meta, positioning, automata}

\usepackage{mathtools}
\DeclarePairedDelimiter{\abs}{\lvert}{\rvert}

\newcommand{\ket}[1]{|#1\rangle}
\newcommand{\ketbra}[2]{\ket{#1}\!\bra{#2}}
\newcommand{\bra}[1]{\langle#1|}

\newcommand{\tr}{\text{tr}}

\def\A{\mathcal{A}}
\def\B{\mathcal{B}}
\def\I{\mathcal{I}}

\newtheorem*{theorem*}{Theorem}

\theoremstyle{definition}
\newtheorem{alg}{\protect\algorithmname}
\theoremstyle{plain}
\newtheorem{defn}{\protect\definitionname}
\theoremstyle{plain}

\theoremstyle{plain}
\newtheorem{prop}{\protect\propositionname}
\theoremstyle{plain}

\theoremstyle{plain}

\theoremstyle{plain}

\providecommand{\conjecturename}{Conjecture}
\providecommand{\definitionname}{Definition}
\providecommand{\lemmaname}{Lemma}
\providecommand{\corollaryname}{Corollary}
\providecommand{\theoremname}{Theorem}
\providecommand{\propositionname}{Proposition}
\providecommand{\algorithmname}{Algorithm}

\newtheorem{result}{Result}
\newtheorem{observation}{Observation}

\def\A{\mathcal{A}}
\def\B{\mathcal{B}}
\def\O{\mathcal{O}}

\def\I{\mathcal{I}}
\def\J{\mathcal{J}}

\def\S{\mathcal{S}}
\def\R{\mathcal{R}}

\def\T{\mathcal{T}}
\def\X{\mathcal{X}}
\def\Y{\mathcal{Y}}

\newcommand{\bk}[2]{\langle#1|#2\rangle}

\newcommand{\braket}[2]{\langle #1 \vert #2 \rangle}

\newcommand{\ema}{$\epsilon$-machine}
\newcommand{\emas}{$\epsilon$-machines}
\newcommand{\etr}{$\epsilon$-transducer}
\newcommand{\etrs}{$\epsilon$-transducers}

\newcommand{\ee}{\mathbf{E}}

\newcommand{\tightoverset}[2]{%
	\mathop{#2}\limits^{\vbox to -.5ex{\kern-0.75ex\hbox{$#1$}\vss}}}

\newcommand{\er}{\sim_\epsilon}
\newcommand{\prob}{\mathrm{Pr}}

\def\tr{\mbox{tr}}

\input{oharp}

\newcommand{\past}[1]{\olharp{#1}}
\newcommand{\future}[1]{\orharp{#1}}

\begin{document}

\title{How Quantum Agents Can Change Which Strategies Are More Complex}
	
\author{Spiros Kechrimparis}
\email{skechrimparis@gmail.com}
\affiliation{School of Computational Sciences, Korea Institute for Advanced Study, Seoul 02455, South Korea}
\affiliation{Nanyang Quantum Hub, School of Physical and Mathematical Sciences, Nanyang Technological University, 637371, Singapore.}

\author{Nix Barnett}
\email{nxbrnt@gmail.com}
\noaffiliation

\author{Mile Gu}
\email{mgu@quantumcomplexity.org}
\affiliation{Nanyang Quantum Hub, School of Physical and Mathematical Sciences, Nanyang Technological University, 637371, Singapore.}
\affiliation{Centre for Quantum Technologies, Nanyang Technological University, 639798, Singapore.}
\affiliation{MajuLab, CNRS-UNS-NUS-NTU International Joint Research Unit, UMI 3654, 117543, Singapore.}

\author{Hyukjoon Kwon}
\email{hjkwon@kias.re.kr}
\affiliation{School of Computational Sciences, Korea Institute for Advanced Study, Seoul 02455, South Korea}
\affiliation{Quantum Universe Center, Korea Institute for Advanced Study, Seoul 02455, South Korea}

 \begin{abstract} Whether winning blackjack or navigating busy streets, achieving desired outcomes requires agents to execute adaptive strategies, strategies where actions depend contextually on past events. In complexity science, this motivates memory as an operational quantifier of complexity: given two strategies, the more complex one demands the agent to track more about the past. Here, we show that conclusions about complexity fundamentally depend on whether agents can process and store quantum information. Thus, while classical agents might find Strategy $\A$ more complex to execute than Strategy $\B$, quantum agents can reach the opposite conclusion. We derive sufficient conditions for such contradictory conclusions and illustrate the phenomenon across multiple scenarios. As a byproduct, our results yield an information-theoretic lower bound on the minimal memory required by any agent - classical or quantum - to execute a given strategy.
\end{abstract}
	
\maketitle

\section{Introduction}

Agents performing complex tasks surround us, from biological organisms adapting to dynamic environments, artificial intelligence engaging in conversation, to the hypothetical Maxwell’s demon that tracks and sorts fast- and slow-moving molecules to reduce entropy. Fundamentally, such autonomous agents operate by (i) interacting with the environment to gather data, (ii) storing relevant data in memory, and (iii) processing this information to decide appropriate future actions. In this way, an agent can execute complex strategies that demand contextual decisions -  decisions that depend not only on what the agent currently sees, but also on what happened many time steps prior. The capability to do so, however, depends crucially on memory. 

Consider a simple example. An agent observes a coin that is flipped repeatedly at discrete time steps. At each time step, the agent is required to announce if the present outcome matches that of the outcome two steps prior. To achieve an optimal success rate, the agent needs memory capable of storing two bits, which is enough to remember the two outcomes from their immediate past. Should we force such an agent to remember a single bit, its performance will invariably degrade. Memory limitations severely constrain which strategies an agent can and cannot execute. 

Consequently, the memory needed to implement a target strategy is a key measure of the strategy's intrinsic complexity \cite{barnett_computational_2015}. Known as statistical complexity, the measure represents the minimal past information any agent executing the strategy must track about its past inputs and outputs. Such a measure then produces a natural hierarchy of what strategies are simple and what strategies are complex. With the advent of quantum computers, we anticipate that agents harnessing quantum memory can execute strategies with less memory~\cite{thompson_using_2017,elliott_quantum_2022}, but we may still expect the \emph{relative order} of which strategies are more complex to remain the same. 

\begin{figure*}[!t]
	\centering
	\includegraphics[trim={4.3cm 4.5cm 6cm 3.8cm},clip,width=1\linewidth]{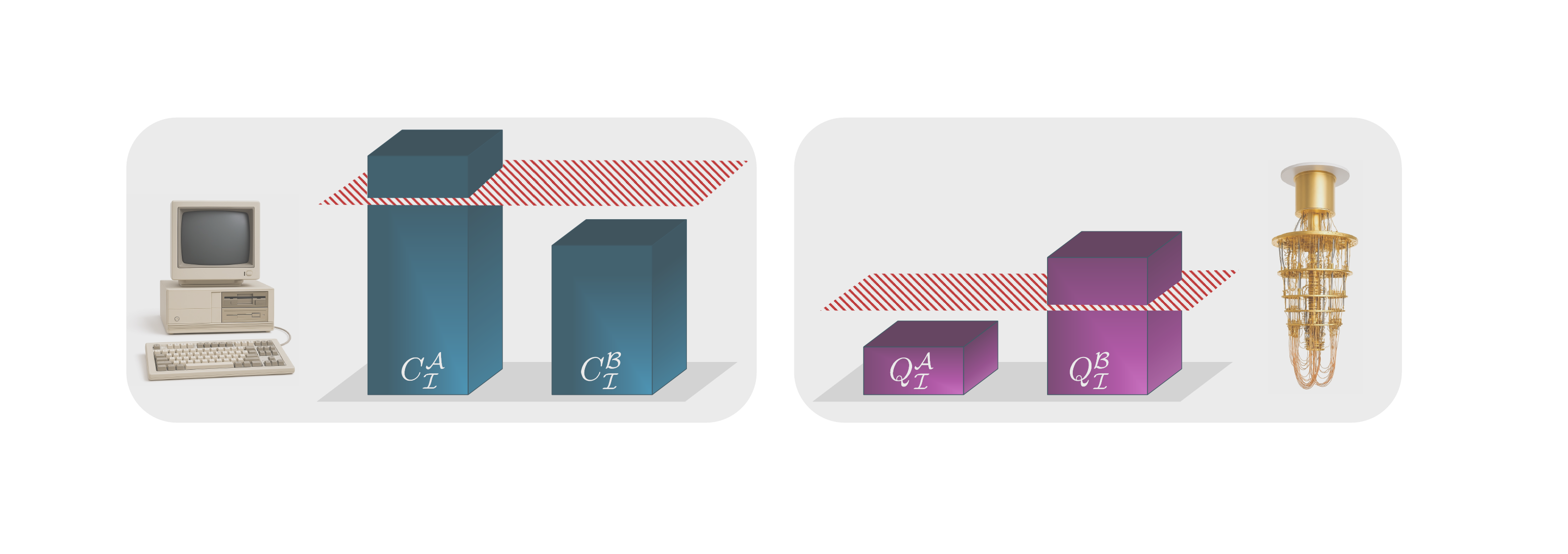}
	\caption{Depiction of a scenario involving an agent who is deciding which of two strategies $\A$ and $\B$ to execute. 
    They may in principle use a device with either a classical or a quantum memory to that effect, each type coming with a different operational memory constraint. When reacting to some input stimuli coming from a stochastic process $\I$, the classical (left) and quantum (right) memory costs $C^{\A}_\I,C^{\B}_\I$ and $Q^{\A}_\I,Q^{\B}_\I$ of executing the two strategies $\A$ and $\B$ by the agent are shown. The red-shaded planes represent the postulated classical (left) and quantum (right) memory constraints. In this visualisation, a classical agent can only execute strategy $\A$, while a quantum agent can only execute strategy $\B$; they arrive at contradictory conclusions on which strategy is harder to implement and thus more complex. }
	\label{Fig: scenario with memory constraints}
\end{figure*}

Here, we illustrate otherwise. We show that when comparing the relative complexity of two strategies, there is a \emph{classical-quantum ambiguity of complexity}. Which strategy requires an agent to track more memory depends crucially on whether they are using classical or quantum memory. Varying a given strategy can make it more complex for classical agents and yet be more easily executed by a quantum agent (see Fig.\@ \ref{Fig: scenario with memory constraints}). Our results generalise a previous finding in predictive modelling of stochastic processes~\cite{aghamohammadiAmbiguitySimplicityQuantum2017,ghafari_errortolerant_2022a} to the case of input-output processes. 
We derive sufficient conditions to detect whether two strategies will exhibit ambiguity of complexity or signal its absence. We demonstrate this phenomenon in various settings, including (i) two agents implementing different strategies while reacting to the same input stimulus, (ii) a single agent implementing a single strategy when reacting to two different input stimuli, and (iii) an agent and their operationally inverse agent implementing the inverse actions. 
Finally, we show that the absence of an absolute relative ordering of strategies is not always reflected in the output actions of the agents when considered alone as stochastic processes. 
As a byproduct, our work also establishes \emph{channel excess entropy} as a non-trivial lower bound on the minimal amount of information any agent, regardless of its underlying physics, must store about the past to execute a given strategy.

The article is structured as follows. Sec.\@ \ref{sec: Framework} introduces our framework and reviews the agentic realisations of stochastic processes and input-output processes, relevant informational theoretic quantities, as well as the current state-of-the-art quantum generalisations. Sec.\@ \ref{sec: main results} describes our main results, including sufficient conditions that can detect the presence or absence of ambiguity of complexity. In Sec.\@ \ref{sec: illustrative examples} we present scenarios and explicit examples that exhibit such ambiguous ordering between the classical and quantum complexities. Finally, Sec.\@ \ref{sec: outputs vs strategies} shows that this ambiguity is not always reflected in the stochastic processes defined by their outputs.

\section{Framework}
\label{sec: Framework}

 \subsection{Agents}
\begin{figure*}[!t]
	\centering
	\includegraphics[trim={12.5cm 3cm 12.2cm 3cm},clip,width=0.9\linewidth]{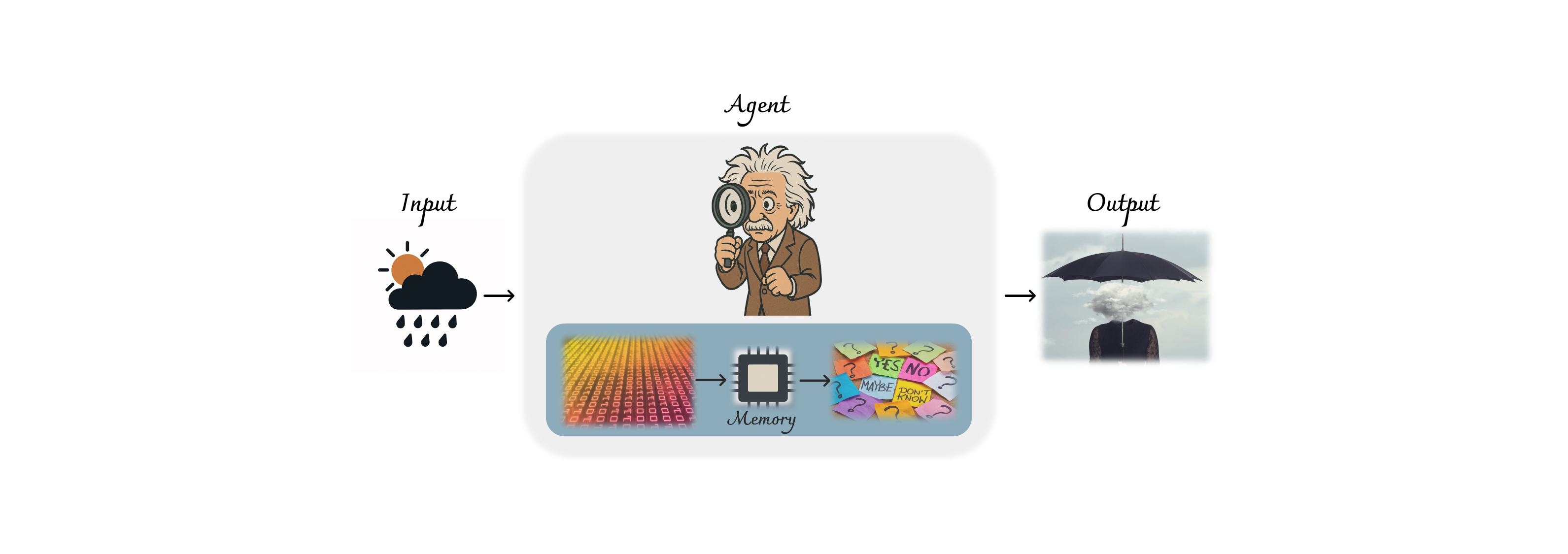}
	\caption{Visual representation of an agent receiving information from their environment, storing the information to a memory, and outputting appropriate actions after processing the stored information according to a desired strategy.}
	\label{Fig: agent}
\end{figure*}

 In a minimal definition, an agent is a system interacting with their environment, recording received stimuli, and outputting appropriate actions in order to implement desired strategies (see Fig.\@ \ref{Fig: agent}). \emph{Computational Mechanics} formalises these concepts. The inputs the agent receives and their output actions can often be taken to be from a \emph{discrete stochastic process} and the \emph{adaptive strategies} that an agent implements are described as \emph{input-output processes}. In this framework, agents are thus automatons that interact with an environment at discrete time steps~\cite{barnett_computational_2015,thompson_using_2017,elliott_quantum_2022}. An input stimulus $x_t$ received from the environment at time $t$ leads to a reaction of the agent, who outputs an action $y_t$, both described by random variables $X_t$ and $Y_t$ taking values from some discrete alphabets $\mathcal{X}$ and $\mathcal{Y}$. Thus, from a mathematical point of view, the streams of inputs and outputs of the agents define stochastic processes, while the agents themselves are implementing input-output processes or channels \cite{barnett_computational_2015}, also referred to as \emph{adaptive strategies}\cite{elliott_quantum_2022}. To optimally characterise the minimum information an agent needs to track, in general, we need to consider semi-infinite histories of joint inputs and outputs (see next section), which we denote by $\olharp{(X,Y)}$.   
We use the following definition of adaptive agents \cite{elliott_quantum_2022}:
\begin{defn}
     An adaptive agent is defined by the tuple $(\X,\Y,S,\epsilon,\Lambda)$, where
     \begin{enumerate}[i)]
         \item $\X$ is the set of input stimuli the agent can recognise,
         \item $\Y$ is the set of output actions the agent can perform,
         \item $\S=\{s_i\}$ is the set of memory states in their memory system, labeled by some index $i=1,\ldots,N$,
         \item $\epsilon:\olharp{(X,Y)}\rightarrow \S$ is an encoding function that determines the memory state to which the agent assigns each history $\olharp{(x,y)}$, 
         \item $\Lambda: \X\times \S \rightarrow \Y \times S$ is the agent's policy, describing how the agent selects action $y$ in response to stimulus $x$ given their current internal state, and how the memory state is updated.
     \end{enumerate}
\end{defn}
The complexity of an agent can then be characterised by how much information the agent must track within its memory - often quantified by the information entropy of $\S$ (see Sec.\@ \ref{sec: framework IO}). Fundamentally, this represents the minimal amount of memory needed to execute each instance of such an agent. As we outline in the following sections, this notion of complexity then leads to a natural definition of complexity for an adaptive strategy. Namely, each strategy may be realised by a plethora of different agents -- its complexity is then given by the complexity of its agentic realisation, when minimised over all such agents \cite{crutchfieldCalculiEmergenceComputation1994, crutchfield_inferring_1989, shalizi_computational_2001, crutchfield_between_2012}. Traditionally, the memory is considered to be classical, and minimisation is defined with that in mind. Quantum mechanics, however, allow $S$ to be quantum mechanical - broadening the class of available agents and thus may lower the complexity of a strategy~\cite{elliott_quantum_2022}. In the following sections, we provide precise mathematical definitions of complexities, as well as other relevant information measures.

\subsection{Classical Modelling of Stochastic Processes}
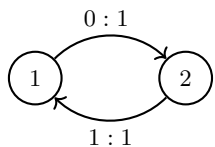
\begin{figure}[!b]
	\centering
	\begin{tikzpicture}[node distance={20mm}, thick, main/.style = {draw, circle,minimum size=0.7cm}] 
	\node[main] (1) {$1$}; 
	\node[main] (2) [ right of=1] {$2$}; 
			
	\draw[->] (1) to [out=45,in=135,looseness=1] node[align=center,midway,above] {$0:1$ } (2); 
	\draw[->] (2) to [out=-135,in=-45,looseness=1] node[align=center,midway,below] {$1:1$}  (1); 
	
	\end{tikzpicture} 
	\caption{The \ema{} of the Period-2 process.}
	\label{Fig: period-2 process ema}
\end{figure}

The inputs and outputs of an agent can be thought of as the emissions of a discrete-time, discrete-alphabet stochastic process; however, they can also be considered to be chosen freely. Such a temporal stochastic process $\olrharp{Y}$ is an infinite sequence $\olrharp{Y}\equiv\ldots Y_{-2}Y_{-1}Y_0 Y_1 Y_2 \ldots$ of discrete random variables $Y_t \,, t\in \mathbb{Z}$ that take values over a countable alphabet $\mathcal{Y}$. By capital symbols, we denote the random variables, while with lowercase symbols, their specific realisations. Finite blocks of outcomes are represented by the sequence of random variables $Y_iY_{i+1}\ldots Y_{j-1}$, denoted by $Y_{i:j}$.
By $\past{Y}=\ldots Y_{-2}Y_{-1}$ and $\future{Y}=Y_0 Y_1 \ldots$ we denote semi-infinite pasts and futures. A stochastic process is defined through its word probabilities $\prob(Y_{t:t+L}) \equiv \{\prob(Y_{t:t+L}=y_{t:t+L})\}$. 
Here we consider stochastic processes that are \emph{stationary}, processes whose word probabilities are invariant under time translation. Another property we require is \emph{ergodicity}, processes whose samples give good estimates of the probabilities. For example, consider a \emph{Biased Coin Process} describing a series of independent coin flips, defined through the word probabilities $\prob(Y_{0:T})=\prob{(Y_0)}\ldots \prob{(Y_{T-1})}$, that is, the process is independent and identically distributed (IID), and $\prob{(Y_t)}=\prob{(Y_\tau)}$ for any $t, \tau$. Another example is a \emph{Period-2 Process} that alternates emissions of 0 and 1 and thus its word probabilities are $\prob(Y_{t:\tau})=\prob{(\ldots 101010\ldots)}=\prob{(\ldots 010101 \ldots)}=\nicefrac{1}{2}$ for any $t,\tau$.
 
Optimal prediction models of stochastic processes are the focus of Computational Mechanics \cite{crutchfieldCalculiEmergenceComputation1994, crutchfield_inferring_1989,shalizi_computational_2001, crutchfield_between_2012}. Here, each instance of a process with a specific past $\past{y}$ has a corresponding conditional future $P(\future{Y}|\past{y})$. The task of prediction is to replicate this conditional future by storing sufficient information about $\past{y}$. The optimal model is the one that stores the minimal amount of information, while still being able to replicate $P(\future{Y}|\past{y})$. This leads to the notion of partitioning pasts into equivalence classes that correspond to identical future predictions, known as \emph{causal states}. The causal states as well as the transition probabilities between them define the so-called \emph{\ema}.

More formally, the \ema{} of a stochastic process is a tuple $(\mathcal{Y};\S; \mathcal{T} )$, where $\mathcal{Y}$ denotes the output alphabet, $\S$ the causal state set, and $\mathcal{T}$ the transition matrix set of the process. The \ema{} is the \emph{unique}, \emph{minimal} and \emph{unifilar}\footnote{Unifilarity is the property that guarantees determinism on the next state of the machine given the current state and emission} presentation of the stochastic process. The causal states essentially represent equivalent classes of pasts that lead to the same futures through an equivalence relation $\er$:
\begin{align}
	\past{y} \er \past{y} {}^\prime \Longleftrightarrow \prob (\future{Y} | \past{Y}=\past{y}) = \prob (\future{Y}  | \past{Y}=\past{y}^\prime) \,. \notag
\end{align}
With the causal states having been obtained the presentation is mathematically given as an \emph{edge-emitting (or Mealy) Hidden Markov Model} \cite{travers_equivalence_2012}. The transitions between causal states are given by a set of \emph{transition matrices}, one for each alphabet symbol,
$\mathcal{T}\equiv \{T^{(y)}\}_{y\in \mathcal{Y}} \,.$
Their elements are explicitly given by 
$T^{(y)} _{ij}= \Pr(Y_{t}=y, S_{t+1}=s_j | S_{t}=s_i) \,,$
representing the probability of transitioning from causal state $s_i$ to state $s_j$ while emitting the symbol $y$. An \ema{} can be visually represented as a graph consisting of nodes and labelled arrows; the nodes correspond to the causal states and the arrows to the transitions between the causal states with labels of the form $y:p$, with $y$ be being the output symbol and $p$ the corresponding probability. For instance, the \ema{} of the aforementioned Period-2 process that deterministically alternates between outputting `0' and `1' is shown in Fig.\@ \ref{Fig: period-2 process ema}.

A quantity of importance is given in the following definition.
\begin{defn}
	The \emph{excess entropy} $\ee$ of a stochastic process is the mutual information between its semi-infinite past, and its semi-infinite future output:
	\begin{align}
		\ee = I[\olharp{Y} ; \orharp{Y}].
	\end{align}
\end{defn}

The excess entropy captures the information from the past contained in the future; it represents the minimal amount of information that needs to be captured by \emph{any} model of a process to correctly reproduce its behaviour.  
As the mutual information between two variables vanishes if the two variables are independent or share no information, it follows that excess entropy vanishes in the case of full randomness or full predictability and is non-zero between these two extreme cases.

In this work, we are interested in the characterisation of complexity or structure. For a stochastic process, this is given by the \emph{statistical complexity}, $C$. Intuitively, the statistical complexity is the amount of past information that needs to be stored in a memory to predict the process' future behaviour. Formally, we have the following definition.
\begin{defn}
	The \emph{statistical complexity} is the Shannon entropy of the distribution over the causal states of the process' \ema{},
	\begin{align}
		C = H[\S]=-\sum_{s_i\in\S}\pi_i \log\pi_i \,,
	\end{align}
	where $\pi_i=\Pr(s_i)$ are the elements of the stationary distribution, the long-term probabilities of the \ema{} being in causal state $s_i$. 
\end{defn}
We remark that the stationary distribution can be directly evaluated as the principal left eigenvector of the total transition matrix $T=\sum_{y}T^{(y)}$.
The statistical complexity is decoupled from the intrinsic randomness of the process; the latter is captured by the entropy rate \cite{ellison_prediction_2009,barnett_computational_2015}, the amount of information or randomness generated per emission.

Importantly, the excess entropy bounds from below the statistical complexity, with equality achieved for reversible processes \cite{guQuantumMechanicsCan2012}.
\begin{prop}
\label{prop: E<=C SP}
    The excess entropy $\ee$ of a stochastic process lower bounds its statistical complexity $C$, $\ee \leq C$.
\end{prop}
A proof of this statement was first given in Ref.\@ \cite{shalizi_computational_2001} and we give a variant of it in Appendix \ref{App: proof of EE and C for passive SP}.
Thus, from the point of view of modelling, the last inequality sets an information-theoretic lower bound on the complexity that a model needs to capture to accurately reproduce the process. As the bound is not, in general, saturated even by the provably optimal classical models, we see that there is a fundamental overhead in classical modelling.

\subsection{Classical Modelling of Input-Output Processes}
\label{sec: framework IO}
    
So far we have only discussed the optimal modelling of stochastic processes, corresponding to agents that emit the same output behaviours regardless of input. In this work, we are interested in agents executing adaptive strategies, where output behaviour at each-time step depends on past inputs. From a mathematical point of view, adaptive strategies that agents execute can be described as \emph{channels} or \emph{input-output processes}. A channel is a coupling between two stochastic processes $\X$ and $\Y$, where one is viewed as the input and the other as the output, and their optimal modelling has been considered in Ref.\@ \cite{barnett_computational_2015}. 
Specifically, we have the following definition.
\begin{defn} \cite{barnett_computational_2015}
	A \emph{channel} or \emph{input-output process} $\olrharp{Y}|\olrharp{X}$ with input alphabet $\X$ and output alphabet $\Y$ is a collection of stochastic processes over alphabet $\Y$, where each process $\olrharp{Y}|\olrharp{x}$ corresponds to a bi-infinite input sequence in $\olrharp{\X}$:
	\begin{align}
		\olrharp{Y}|\olrharp{X} \equiv \{\olrharp{Y}|\past{x}\} _{\olrharp{x}\in \olrharp{\X}}.
	\end{align}
	In other words, each fixed realisation $\olrharp{x}=\ldots x_{-1}x_0 x_1 \ldots$ over the input alphabet $\X$ is mapped to a stochastic process $\olrharp{Y}| \olrharp{x} $ over alphabet $\Y$.
\end{defn}

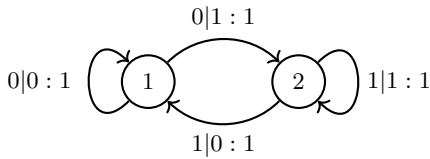
\begin{figure}[!t]
	\centering
		\begin{tikzpicture}[node distance={20mm}, thick, main/.style = {draw, circle,minimum size=0.7cm}] 
					\node[main] (1) {$1$}; 
					\node[main] (2) [ right of=1] {$2$}; 
					
					\draw[->] (1) to [out=45,in=135,looseness=1] node[align=center,midway,above] {$0|1:1$} (2); 
					\draw[->] (2) to [out=-135,in=-45,looseness=1] node[align=center,midway,below] {$1|0:1$}  (1); 			\draw[->] (1) to [out=225,in=135,looseness=5] node[align=center,midway,left] {$0|0:1$ } (1);
					\draw[->] (2) to [out=45,in=315,looseness=5] node[align=center,midway,right] {$1|1:1$ } (2);
				\end{tikzpicture} 
			\caption{A one-step delay channel.}
            \label{Fig: delay channel}
\end{figure}

In analogy to the case of stochastic processes, by defining an appropriate equivalence relation over joint input and output pasts, one obtains the channel's causal states. Specifically, we have the equivalence relation $\er$ over joint pasts $\olharp{z}=\olharp{(x,y)}$:
\begin{align}
	&\olharp{(x,y)} \er \olharp{(x,y)}^\prime \,\,\,\, \Longleftrightarrow \notag \\
    &\,\,\,\, \Pr\left(\orharp{Y}| \orharp{X} ,\olharp{(X,Y)}=\olharp{(x,y)}\right)  \notag \\
    &\,\,\,\,\,\,\,\,=  \Pr\left(\orharp{Y}| \orharp{X} ,\olharp{(X,Y)}=\olharp{(x,y)}^\prime\right) \,,
\end{align}
that partitions the set of joint pasts $\olharp{\mathcal{Z}}=\olharp{(\X,\Y)}$ of all input-output pasts into equivalent classes. These classes are the \emph{causal states} of the input-output process denoted by $\S$. 
Transitions between causal states are described by transition matrices
$\mathcal{T}\equiv \{T^{(y|x)}\}_{x\in \mathcal{X},y\in \mathcal{Y}} \,,$
with elements
$T^{(y|x)} _{ij}= \Pr(Y_{t}=y, S_{t+1}=s_j |  X_t =x, S_{t}=s_i) \,.$
Note that the transition matrices are \emph{input-dependent}, as are all informational quantities we are interested in. 
\begin{defn}
	The \emph{\etr{}} of an input-output process is defined as the tuple $(\mathcal{X},\mathcal{Y},\mathcal{S},\mathcal{T})$ consisting of the input and output alphabets, the set of causal states, and the conditional transition probabilities.
\end{defn}
The \etr{} is the unique and maximally predictive, unifilar \footnote{Unifilarity of an \etr{} is the property that guarantees determinism on the next state of the machine given the current state, as well as current input and emission \cite{barnett_computational_2015}} presentation of a channel that minimises the statistical complexity \cite{barnett_computational_2015}. As with \emas{}, \etrs{} can be visually represented as graphs consisting of nodes and labelled arrows; the nodes correspond to the causal states and the arrows to the transitions between the causal states with labels of the form $y|x:p$, where $x$ and $y$ are the input and output symbols, respectively, and $p$ the corresponding probability. For instance, a \emph{one-step delay channel}, which at each time step outputs the input at the previous time step, is shown in Fig.\@ \ref{Fig: delay channel}.

An \emph{input-dependent statistical complexity} of the \etr{} is defined as the Shannon entropy of the stationary distribution, $\pi_\I$, over the causal states of the \etr, where the index $\I$ denotes the fact that the distribution depends on the input process $\I = \olrharp{X}$. In general, different input processes will drive the \etr{} to different stationary distributions over its causal states. Specifically, we have:
\begin{defn}
	The \emph{input-dependent statistical complexity} $C_\I$ of an \etr{}, when driven by an input-process $\I=\olrharp{X}$, is given by 
	\begin{align}
		C_\I=H(\pi_\I) = -\sum_{i= 1}^{|\S|} \pi_\I (s_i) \log \pi_\I (s_i)\,,
	\end{align}
where $\pi_\I (s_i)$ denotes the probability of being in causal state $s_i$.
\end{defn}

 Finally, one can define an input-independent quantity by taking the supremum over all possible input stochastic processes, obtaining the \emph{channel complexity}, $\overline{C}=\sup _X C_X$, which gives an upper bound on the complexity of the \etr. A further upper bound is given by the \emph{topological complexity}, $C_0$, obtained as the Shannon entropy of the uniform distribution over the causal states of the \etr. However, not all \etrs{} can be driven to a uniform distribution over causal states, and thus this bound is not always attainable. In this work, we will not use the input-independent quantities as we are interested in the behaviour of the complexities of agents when driven by a specific given input stimulus.

\subsection{Quantum Modelling of Stochastic Processes and Input-Output Processes}

\begin{figure*}[!t]
	\centering
	\includegraphics[trim={0cm 0cm 0cm 0cm},clip, width=0.7\linewidth]{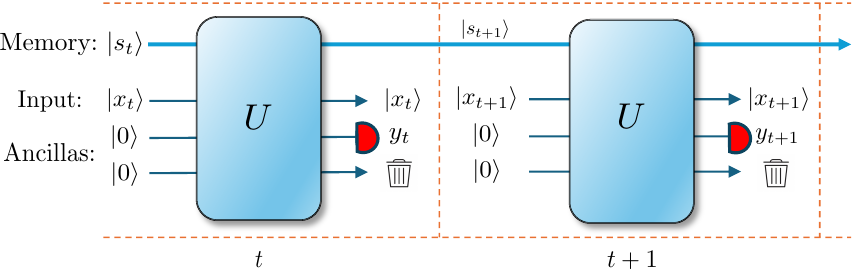}
	\caption{Circuit representation of the implementation of a quantum agent (see main text). }
	\label{Fig:circuit}
\end{figure*} 

Recently there has been interest in constructing quantum models of stochastic processes and exploring potential advantages. Several results have emerged: from exhibiting memory advantages for both stochastic \cite{guQuantumMechanicsCan2012,suen_classical-quantum_2017, garner_provably_2017, ghafari_dimensional_2019,elliott_extreme_2020}, and input-output processes \cite{thompson_using_2017,elliott_quantum_2022}, to causal asymmetry \cite{thompson_causal_2018,kechrimparis_causal_2023}.

Quantum model design assigns the causal states of the optimal classical model to states of a quantum system, along with a quantum evolution and a measurement protocol to extract the classical outputs. A quantum model is said to be faithful if the output conditional word probabilities are indistinguishable from those of the classical one. 
In contrast to the classical case, proving the optimality of a candidate quantum model among all quantum models is highly nontrivial. The crux of the problem is to choose the quantum causal states so that to minimise the von Neumann entropy while remaining statistically faithful. Given the intricate behaviour of the von Neumann entropy \cite{jozsa_distinguishability_2000}, optimality has only been proven in a few specific cases.

Nevertheless, there exist systematic procedures to construct quantum models that perform better than any classical models in the event that optimal classical models still exhibit causal waste. This occurs whenever the process at each time step is \emph{logically irreversible} \cite{guQuantumMechanicsCan2012}: if there exist two distinct states of the \ema{} such that upon emitting the same symbol, they transition with nonzero probabilities to a coinciding state.
Analogously, a quantum model of an input-output process exhibits memory advantage whenever the process is \emph{step-wise inefficient}\cite{thompson_using_2017}: there exist two distinct causal states of the \etr{} such that for any input symbol there is non-zero probability that they transition to the same causal state while emitting a coinciding output.

We briefly review quantum models of input-output processes that guarantee improvement over classical models whenever the \etr{} is step-wise inefficient. Note that the case of stochastic processes is trivially included and corresponds to an input alphabet of exactly one symbol; thus, in the following, the case of passive stochastic processes is implicitly considered.  It was previously shown that quantum agents can be optimised by considering only those obeying the following constraints~\cite{elliott_quantum_2022}:
\begin{enumerate}[(i)]
    \item The agent receives input stimuli $x_t$ encoded in the computational basis states ${\ket{x_t}}$.
    \item The input stimulus is not consumed by the evolution of the agent; $\Lambda$ preserves the input tape.
    \item The agent delivers output actions $y_t$ via projective measurements in the computational basis states ${\ket{y_t}}$.
    \item The memory states are pure and in one-to-one correspondence with the strategy's causal states $\S$.
\end{enumerate}

For instance, given an \etr{} with conditional transition probabilities $T_{ij}^{(y|x)}$, a particular encoding of causal states into quantum states \cite{thompson_using_2017} is given by
\begin{align}
	\ket{s_i}=\otimes_x \ket{s_i^{x}}\,, \label{eq:quantum encoding 1}
\end{align}
with  
\begin{align}
    \ket{s_i^x}=\sum_{k}^{\abs{\S}}\sum_{y}^{\abs{\Y}} \sqrt{T_{ik}^{(y|x)}} \ket{y}\ket{k}\,, \label{eq:quantum encoding 2}
\end{align}
where summation over index $k$ runs over the number of causal states (rows or columns of the transition matrices) and summation over $y$ over the output symbols; $\ket{y}$ denotes an orthonormal basis in a Hilbert space of dimension $\abs{\Y}$ while $\ket{k}$ an orthonormal basis of dimension $\abs{\S}$; $\otimes_x$ denotes the direct product over input symbols, $x\in\X$. Note that we have dropped the $t$ subscript to not overload notation. A unitary is then acting to the encoded causal states plus ancillas followed by an appropriate measurement that extracts correctly the emissions of the process; in addition, the measurement also prepares the quantum state of the machine to the state that corresponds to the correct next causal state of the \etr. The unitary is implicitly defined through the equation
\begin{align}
	U \ket{s_i} \ket{x} \ket{0} \ket{0} = \sum_y \sqrt{T_{ij}^{(y|x)}} \ket{s_j} \ket{x} \ket{y} \ket{ \psi},
\end{align}
where $\ket{x}$ denotes the input stimulus encoded as a quantum state and $\ket{\psi}$ `junk' states that are being discarded.
A schematic representation is shown in Fig. \ref{Fig:circuit}, with the details of the encoding given in Ref.\@ \cite{thompson_using_2017}, and a better approach introduced in Ref.\@ \cite{elliott_quantum_2022}.

As the process is repeated, the memory state is being updated accordingly, leading to the average state of the quantum memory,
\begin{align}
    \rho=\sum_{i=1}^{|\S|} \pi_\I (s_i) \ketbra{s_i}{s_i} \,,
\end{align}
when driven by an input process $\I = \olrharp{X}$, and where $\pi_\I (s_i)$
is the long-term probability of being in causal state $s_i$.
\begin{defn}
    The quantum complexity of a quantum agent is the von Neumann entropy of the average state of the quantum memory,
\begin{align}
	Q_\I = -\tr (\rho \log \rho)\,,
\end{align} 
when driven by an input process $\I = \olrharp{X}$. 
\end{defn}
It is clear that if the quantum states of the encoding are orthogonal, the complexity reduces to the classical complexity, showcasing the source of the quantum improvement. Thus, we generally have $Q_\I \leq C_\I$. We note that the same conclusion holds for passive stochastic processes, since a trivial input, i.e. an input process with an alphabet with a single symbol, reduces the case of input-output processes to the case of passive stochastic processes. 

Returning to the case of passive stochastic processes, even though the quantum complexity of an irreversible process is lower than that of the classical one, it is still lower bounded by the excess entropy.
\begin{prop}
\label{prop: E<=Q SP}
    The quantum complexity $Q$ of a quantum model is lower bounded by the excess entropy $\ee$, $\ee \leq Q$.
\end{prop}
This was first pointed out in Ref.\@ \cite{guQuantumMechanicsCan2012} but no proof was provided. Although we give the full proof in Appendix \ref{App: proof of EE and Q for passive SP}, we sketch it here. 
	A classical model maps pasts into causal states. Moreover, a quantum model consists of an encoding function that maps classical states of the \ema{} to quantum states, as well as a quantum evolution that extracts the classical emissions. 
    By iteration, the entire future can be generated. As the excess entropy is the mutual information between past and future classical emissions, the result follows from Holevo's bound \cite{holevo_bounds_1973, nielsen_quantum_2010}.

\section{Classical-Quantum Ambiguity of Complexity of strategies}
\label{sec: main results}

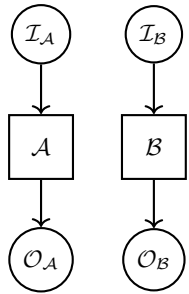
\begin{figure}[!t]
	
	\centering
	\begin{tikzpicture}[node distance=1.5cm, thick,square/.style={regular polygon,regular polygon sides=4,minimum size=1.2cm}, main/.style={draw, circle,minimum size=0.7cm}]
		\node[main] (1) {$\I_\A$}; 
		\node [main] [right of=1] (2) {$\I_\B$};
		\node[square, draw] [below of=1] (3) {$\A$}; 
		\node[square, draw] [right of=3] (4) {$\B$};
		\node[circle, draw]  [below of=3] (5) {$\O_\A$}; 
		\node[circle, draw] [right of=5] (6) {$\O_\B$};
		\draw[->] (1) -- (3);
		\draw[->] (2) -- (4);
		\draw[->] (3) --(5);
		\draw[->] (4) --(6);
	\end{tikzpicture} 
	\caption{The general setting involving two agents implementing strategies $\A$ and $\B$, while reacting to stimuli $\I_\A$ and $\I_\B$, respectively.}
	\label{Fig:general setting}	
\end{figure}
We now show that scenarios involving agents implementing strategies while reacting to input stimuli can lead to different conclusions regarding the complexity of the behaviour they exhibit, depending on whether they employ classical or quantum memories. Specifically, we consider two agents executing two input-output processes $\A$ and $\B$, while reacting to stimuli from some input processes $\I_\A$ and $\I_\B$, respectively. When driven by these, the agents will emit outputs that define stochastic processes $\O_\A$ and $\O_\B$, respectively. This general setting is shown pictorially in Fig.\@ \ref{Fig:general setting}. By $C_\I ^{\A}, C_\I ^{\B}$ and $Q_\I ^{\B}, Q_\I ^{\B}$ we denote the classical and quantum complexities of $\A$ and $\B$ when driven by inputs $\I_\A$ and $\I_\B$ respectively. Then, we have in general that the quantum complexities can be lower than the classical ones, that is,  $C_\I ^{\A}\geq Q_\I ^{\A}$ and $C_\I ^{\B}\geq Q_\I ^{\B}$. Moreover if, say, $C_\I ^{\A}> C_\I ^{\B}$, which implies that agent $\A$ exhibits more complex behaviour according to a classical description, it is natural to expect a consistently assigned ordering by a quantum description, that is, $Q_\I ^{\A}> Q_\I ^{\B}$. In other words, naturally one would expect classical and quantum models of the two strategies to arrive at similar conclusions regarding which agent's behaviour is more complex. However, we show that it is possible to arrive at contradictory conclusions.
\begin{result}
    There is a classical-quantum ambiguity in the complexities of adaptive strategies executed by agents.
\end{result}

We first derive mathematical conditions that indicate the presence of such an inconsistency and, in the following section, we consider illustrative scenarios and specific examples where this phenomenon manifests. 

Our approach involves first deriving an analogue of the excess entropy for an input-output process - which then bounds the minimal memory of any agent executing this process. Let $\I=\olrharp{X}$ denote the input process to which an agent implementing an input-output process $\A$ is reacting, and by $\O=\olrharp{Y}$ the stochastic process defined by the agent's output actions. We also denote by $\J=\olrharp{(X,Y)}$ the stochastic process of joint inputs and outputs \cite{barnett_computational_2015}. We remark that in the following, our notation is such that informational quantities which have input-dependence, such as those of input-output processes, will have a subscript signaling this fact.     
\begin{defn}
	The \emph{channel excess entropy} $\ee^\A_\I$ from an input process $\I$ to an output process $\O$ is the mutual information between a channel's infinite bivariate past, and its infinite future output given infinite future input:
	\begin{align}
		\ee^\A_\I = I[\olharp{(X,Y)} ; \orharp{Y}|\orharp{X}].
	\end{align}
\end{defn}

In analogy to the case of \emas{}, the channel excess entropy captures the minimal amount of information that must be stored by a model of an input-output process in order to correctly reproduce it, given input. 

We now show that the channel excess entropy lower bounds the classical channel statistical complexity.

\begin{prop}
\label{prop: E<=C IOP}
    The channel statistical complexity $C^{\A}_{\I}$ of a channel $\A$ given an input $\I$ is lower bounded by the channel's excess entropy $\ee^{\A}_{\I}$, $\ee^{\A}_{\I} \leq C^{\A}_{\I}$.
\end{prop}
The proof leverages a conditional version of the data-processing inequality \cite{cover_elements_2005}, with full details given in Appendix \ref{App: proof of EE and C for passive IP}. Next, we show that the quantum complexity is also lower bounded by the channel excess entropy (see proof in Appendix\@ \ref{App: proof of EE and Q for passive IP}:

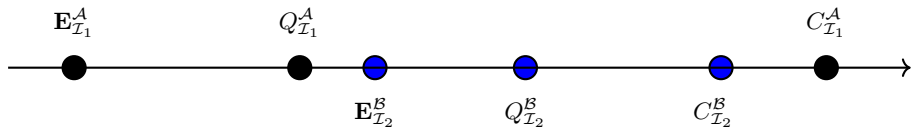
\begin{figure*}[!t]
	\centering
	\begin{tikzpicture}[node distance={1mm}, thick, main/.style = {draw, circle,minimum size=0.2cm,fill=black}, main2/.style = {draw, circle,minimum size=0.2cm,fill=blue}] 
		\node (1) {}; 
		\node (2) [right=12cm of 1]{};
		
		\node[main] at (1, 0)    {};
		\node[main] at (4, 0)    {};
		\node[main] at (11, 0)   {};
		\node[] at (1, 0.6)  {$\ee^\A_{\I_1}$};
		\node[] at (4, 0.6)  {$Q^\A_{\I_1}$};
		\node[] at (11, 0.6)  {$C^{\A}_{\I_1}$};	
		
		\node[main2] at (5, 0)   {};
		\node[main2] at (7, 0)    {};
		\node[main2] at (9.6, 0)   {};
		\node[] at (5, -0.6)  {$\ee^\B_{\I_2}$};
		\node[] at (7, -0.6)  {$Q^\B_{\I_2}$};
		\node[] at (9.5, -0.6)  {$C^\B_{\I_2}$};	
		\draw[->] (1) -- (2);
	\end{tikzpicture} 
	\caption{A visualization of the sufficient condition for ambiguous ordering of the complexities. In black we show the excess entropy and complexities of $\A$, while with blue those of $\B$.}
	\label{Fig:ambiguous condition}	
\end{figure*}

\begin{prop}
\label{prop: E<=Q IOP}
    The quantum complexity $Q^\A_\I$ of a quantum model is lower bounded by the channel excess entropy $\ee^\A_\I$, $\ee^\A_\I \leq Q^\A_\I$.
\end{prop}

We remark that, in general, a channel can be such that allows influences to the present from the future, meaning that one can consider input-output processes such that future inputs can, in principle, influence the present behaviour. Although mathematically possible, this does not correspond to expected physical behaviour where causality is a standard assumption \cite{barnett_computational_2015}. 
As a result, in this work, we only consider \emph{causal} or \emph{anticipation-free} channels.
\begin{defn}
    A \emph{causal channel} or \emph{anticipation-free} channel satisfies
    \begin{align}
        \Pr(Y_{t:t+L}|\olrharp{X})=\Pr(Y_{t:t+L}|\olharp{X}_{t+L})\,.
    \end{align}
    That is, the channel is completely characterised by its behaviour on semi-infinite pasts.
\end{defn}

The meaning of the last definition is clear: the conditional probabilities of future output sequences of any length depend only on the past and present inputs, but not on inputs from the future. For causal channels, we have the following useful decomposition for the excess entropy:
\begin{prop}
\label{prop: decomposition of E for IOP}
	 Given a causal input-output process, its excess entropy $\ee^{\A}_\I$ decomposes in terms of the excess entropies of the joint and input processes, $\ee^{\J}$ and $\ee^{\I}$ respectively, as 
	\begin{align}
		\ee^{\A}_\I = \ee^{\J} -  \ee^{\I}. \label{eq:agent excess entropy}
	\end{align}
\end{prop}
The proof is given in Appendix\@ \ref{App: proof of decomposition of EE} and follows by using the definition of the excess entropies involved and applying known mutual information identities.

We find that the excess entropy decomposes in alignment with the intuition that the minimum memory the agent needs to store equals the minimum memory needed for the process over joint inputs and outputs, in excess of what is provided by the input.
Proposition \ref{prop: decomposition of E for IOP} shows that the channel excess entropy is directly related to the excess entropies of two different stochastic processes, the input process and the stochastic process of over joint inputs and outputs, allowing us to evaluate it with existing methods for \emas{} \cite{ellison_prediction_2009, crutchfield_times_2009}.

We can summarise the last three propositions in the following result:
\begin{result}
The excess entropy of a causal channel $\ee^{\A}_\I$ provides a lower bound on both the classical and quantum complexities of the channel given an input, denoted by $C_\I ^{\A}$ and $Q_\I ^{\A}$. That is, we have
\begin{align}
	\ee^{\A}_\I =  \ee^{\J} - \ee^{\I}  \leq Q^\A_\I \leq C^\A_\I \,.
\end{align}
\end{result}
We can use the last inequality to derive sufficient conditions for ambiguity to occur or not. 
We denote with $C_{\I_\A}^\A$ and $\ee_{\I_\A}^\A$ the classical complexity and excess entropy of $\A$ when driven by $\I_\A$, and similarly with $C_{\I_\B}^\B$ and $\ee_{\I_\B}^\B$ for input-output process $\B$ when driven by $\I_\B$.  Moreover, assume without loss of generality that $C_{\I_\A}^\A>C_{\I_\B}^\B$. We have the following result:
\begin{result}
\label{result: ambiguous ordering}
	Let $\A$ and $\B$ denote two input-output processes driven by inputs from two stochastic processes $\I_\A$ and $\I_\B$, respectively. Let their excess entropies be denoted by $\ee^\A_{\I_\A}, \ee^\B_{\I_\B}$, and their quantum complexities by $Q^\A_{\I_\A}, Q^\B_{\I_\B}$. Moreover, without loss of generality, assume that $C^\A_{\I_\A} > C^\B_{\I_\B}$. Then, if the inequality
	\begin{align}
		\ee^\B_{\I_\B} > Q^\A_{\I_\A} \,, \label{eq:sufficient ambiguous}
	\end{align}
	is satisfied, we have an ambiguous ordering between classical and quantum complexities. That is, we simultaneously have 
	\begin{align}
		C^\A_{\I_\A} > C^\B_{\I_\B}\ \  \text{and} \ \ Q^\A_{\I_\A} < Q^\B_{\I_\B} \,.
	\end{align}
\end{result}
The proof of this statement follows from the definition of the excess entropy and the inequalities connecting it to the classical and quantum complexities. Note that this statement is only a sufficient condition. 
This situation is graphically demonstrated in Fig.\@ \ref{Fig:ambiguous condition}.

We can also make a similar statement for the case with consistent ordering. Specifically,
\begin{result}
\label{result: unambiguous ordering}
	Let $\A$ and $\B$ denote two input-output processes driven by inputs from two stochastic processes $\I_{\A}$ and $\I_{\B}$, respectively. Let their excess entropies be denoted by $\ee^\A_{\I_{\A}}, \ee^\B_{\I_{\B}}$, and their quantum complexities by $Q^\A_{\I_{\A}}, Q^\B_{\I_{\B}}$. Moreover, without loss of generality assume that $C^\A_{\I_{\A}} > C^\B_{\I_{\B}}$. Then, if the inequality
	\begin{align}
		\ee^\A_{\I_{\A}} > Q^\B_{\I_{\B}}
        \label{eq:sufficient unambiguous}
	\end{align}
is satisfied, we have a consistent ordering between classical and quantum complexities. That is, we have 
	\begin{align}
		C^\A_{\I_{\A}} > C^\B_{\I_{\B}}\ \  \text{and} \ \ Q^\A_{\I_{\A}} > Q^\B_{\I_{\B}} \,.
	\end{align}
\end{result}
The last condition is also only sufficient. 

The usefulness of both conditions, Eqs.\@ \eqref{eq:sufficient ambiguous} and \eqref{eq:sufficient unambiguous}, lies in the following fact. Given that it is, in general, a hard problem to obtain the optimal quantum model for a given input-output process, these sufficient conditions can be used to detect the presence or absence of an ambiguous ordering between the classical and quantum complexities, whenever only one of the optimal quantum models for a pair of input-output processes is known. 
In fact, given that the statistical complexity of the optimal model lower bounds the complexity of any quantum model, Results \ref{result: ambiguous ordering} and \ref{result: unambiguous ordering} hold for the quantum complexities of any quantum model on the RHS of Eqs.\@ \eqref{eq:sufficient ambiguous} and \eqref{eq:sufficient unambiguous}, not necessarily an optimal one. 

In the next section we demonstrate the results by considering specific scenarios and explicit examples involving classical and quantum agents that can exhibit classical-quantum ambiguity of complexity.

\section{Illustrative Scenarios}
\label{sec: illustrative examples}
We now demonstrate the results of the last section with different scenarios and particular examples. Let $\I_\A,\I_\B$ be stochastic processes that are taken as inputs of the input-output stochastic processes $\A$ and $\B$, respectively.
 We will show the inconsistency between classical and quantum complexities in three scenarios, which can be seen as special cases of the general setting shown in Fig.\@ \ref{Fig:general setting}. 
 The first consists of two different strategies that are executed by agents reacting to the same environmental stimulus. This corresponds to taking the two inputs $\I_\A$ and $\I_B$ in the general setting shown in Fig.\@ \ref{Fig:general setting} to be the same, i.e., $\I_\A=\I_\B=\I$.
 The second consists of an agent that implements a single strategy when reacting to two different inputs, and corresponds to identifying $\A=\B$. The last scenario is that of an agent implementing a certain strategy and an agent implementing their operational inverse actions. This case corresponds to letting $\I_\A=\O_B$ and $\I_\B=\O_A$. This is because for the agent that is implementing the operational inverse of the first, the role of inputs and outputs is reversed.

\subsection{Two agents reacting to the same stimulus \label{sec: 2 agents 1 stimulus}}
	
\begin{figure}[!b]
		
		\centering
		\begin{tikzpicture}[node distance=1.5cm, thick,square/.style={regular polygon,regular polygon sides=4,minimum size=1.2cm}, main/.style={draw, circle,minimum size=0.8cm}]
			\node[main] (1) {$\I$}; 
			
			\node[square, draw] [below left =0.7cm and 0.3cm of 1] (2) {$\A$}; 
			\node[square, draw] [below right =0.7cm and 0.3cm of 1] (3) {$\B$};
			\node[main]  [below = 1cm of 2] (4) {$\O_\A$}; 
			\node[main] [below = 1cm of 3] (5) {$\O_\B$};
			\draw[->] (1) to [out=180,in=90,looseness=0.8] (2);
			\draw[->] (1) to [out=0,in=90,looseness=0.8] (3);
			\draw[->] (2) --(4);
			\draw[->] (3) --(5);
		\end{tikzpicture} 
		\caption{Two agents executing strategies $\A$ and $\B$ while reacting to the same stimulus.}
		\label{Fig:First scenario}	
\end{figure}
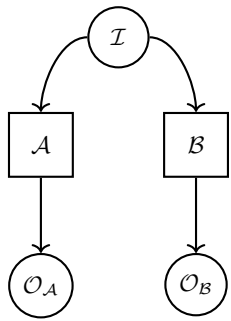

The first scenario we consider consists of two agents implementing two strategies $\A$ and $\B$, driven by the same stimulus $\I$ and outputting actions $\O_\A$ and $\O_\B$, respectively. In its general form, this scenario is depicted in Fig.\@ \ref{Fig:First scenario}.
 We will show the following.

\begin{result}
\label{result: ambiguous ordering for two IO}
    There exist pairs of input-output processes $\A$ and $\B$ and an input process $\I$ such that that the difference of their classical complexities $C_\I^\A$ and $C_\I^\B$ when driven by input $\I$ is positive, $C_\I^\A - C_\I^\B >0$,  while the difference of their quantum complexities $Q_\I^\A$ and $Q_\I^\B$ is negative, $Q_\I^\A - Q_\I^\B <0$.
\end{result}

As a concrete example, let us consider that the two agents, Alice and Bob, wish to simulate the behaviour of two particle detectors. Here we consider input and output alphabets consisting of the same two symbols $\X,\Y=\{0,1\}$. Symbol 0 (1) should be interpreted to mean that the particle was not (was) present.

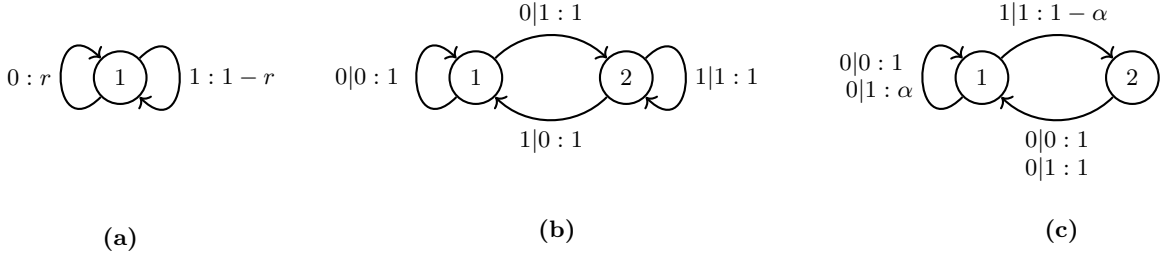
\begin{figure*}[!t]
	\centering

        \begin{tikzpicture}[node distance={20mm}, thick, main/.style = {draw, circle,minimum size=0.7cm}] 
		\node[main] (1) {$1$};
        \node[] (a) [below = 15mm of 1] {\bf (a)};
		\draw[->] (1) to [out=225,in=135,looseness=5] node[align=center,midway,left] {$0: r $} (1);
		\draw[->] (1) to [out=45,in=-45,looseness=5] node[align=center,midway,right] {$1:1-r$ } (1);

					\node[main] (1a) [ right=40mm of 1] {$1$}; 
                    \node[] (ba) [below right = 15mm and 4.5mm of 1a] {\bf (b)};
					\node[main] (2a) [ right of=1a] {$2$}; 
					
					\draw[->] (1a) to [out=45,in=135,looseness=1] node[align=center,midway,above] {$0|1:1$} (2a); 
					\draw[->] (2a) to [out=-135,in=-45,looseness=1] node[align=center,midway,below] {$1|0:1$}  (1a); 			\draw[->] (1a) to [out=225,in=135,looseness=5] node[align=center,midway,left] {$0|0:1$ } (1a);
					\draw[->] (2a) to [out=45,in=315,looseness=5] node[align=center,midway,right] {$1|1:1$ } (2a);

					\node[main] (1c) [ right=60mm of 1a] {$1$}; 
                    \node[] (cc) [below right =15mm and 4.5mm of 1c] {\bf (c)};
					\node[main] (2c) [ right of=1c] {$2$}; 
					
					\draw[->] (1c) to [out=45,in=135,looseness=1] node[align=center,midway,above] {$1|1:1-\alpha$ } (2c); 
					\draw[->] (2c) to [out=-135,in=-45,looseness=1] node[align=center,midway,below] {$0|0:1$\\$0|1:1$}  (1c); 
					\draw[->] (1c) to [out=225,in=135,looseness=5] node[align=center,midway,left] {$0|0:1$ \\$\,\,\,\,0|1:\alpha$  } (1c);		
	\end{tikzpicture}
	
			\caption{(a) The Biased Coin process. (b) Alice is implementing a noiseless detector with delay. (c) Bob is implementing a noisy detector with dead time. 
            }
            \label{Fig: Alice Delay channel and Bob}  
\end{figure*}

Alice is simulating a noiseless detector $\mathcal{A}$ with one time step \emph{delay}: the detector stores the information about the presence or absence of the particle at the current time step and announces it as its output at the next time step. In order to successfully implement the behaviour of $\A$, Alice needs to store 1 bit, her memory needs to have two states, and the corresponding \etr{} is shown in Fig.\@ \ref{Fig: Alice Delay channel and Bob}.

On the other hand, Bob is simulating the behaviour of a \emph{noisy detector with dead time} \cite{knoll_radiation_2010,migdall_singlephoton_2013}: the presence of an input stimulus 1 (presence of the particle) is detected correctly with probability $1-\alpha$, followed by a one time step relaxation period, during which the device is reset to its normal state and no detection is possible.
More specifically, detector $\mathcal{B}$ correctly identifies the absence of the signal and emits 0.
In the presence of a particle, however, it introduces two different types of noise. Specifically, (i) if $\B$ is in its first state, with probability $\alpha$ it will incorrectly detect a 0 (no presence of a particle) and remain in the first state, otherwise correctly output a 1 (detection of particle) and transition to the second state, (ii) while if $\B$ is in its second state, i.e. detected a particle in the previous time step, it completely ignores the input and always outputs a 0 (no particle detection) while resetting back to the first state. In Fig.\@ \ref{Fig: Alice Delay channel and Bob} we depict the \etr{} that consists of two causal states and two-symbol input and output alphabets.

Let us also assume that the original series of input stimuli $\I$ is generated by a simple IID process, the Biased Coin process, which consists of a series of biased coin flips with some bias $r$, shown in Fig.\@ \ref{Fig: Alice Delay channel and Bob}. That is, a particle is randomly generated (or not) at each time step with some probability $1-r$ (or $r$). 

For this input, the statistical complexity of the agents can be readily calculated. Alice's and Bob's classical complexities are found to be (see Appendix \ref{app: 2 agents 1 stimulus}) 
\begin{align}
	C_\I^\A= h(r) \qquad C_\I^\B = h(b),
\end{align}
where $h(q) = -q \log q-(1-q)\log(1-q)$ denotes the binary entropy and $b=\nicefrac{1}{(1+(1-r)(1-\alpha))}$. Note that the complexity of Bob depends on both the bias $r$ of the input as well as the parameter $\alpha$.

 \begin{figure*}[!t]
	\centering
	\includegraphics[width=0.9\linewidth,trim={0cm 1.7cm 0cm 1.8cm},clip]{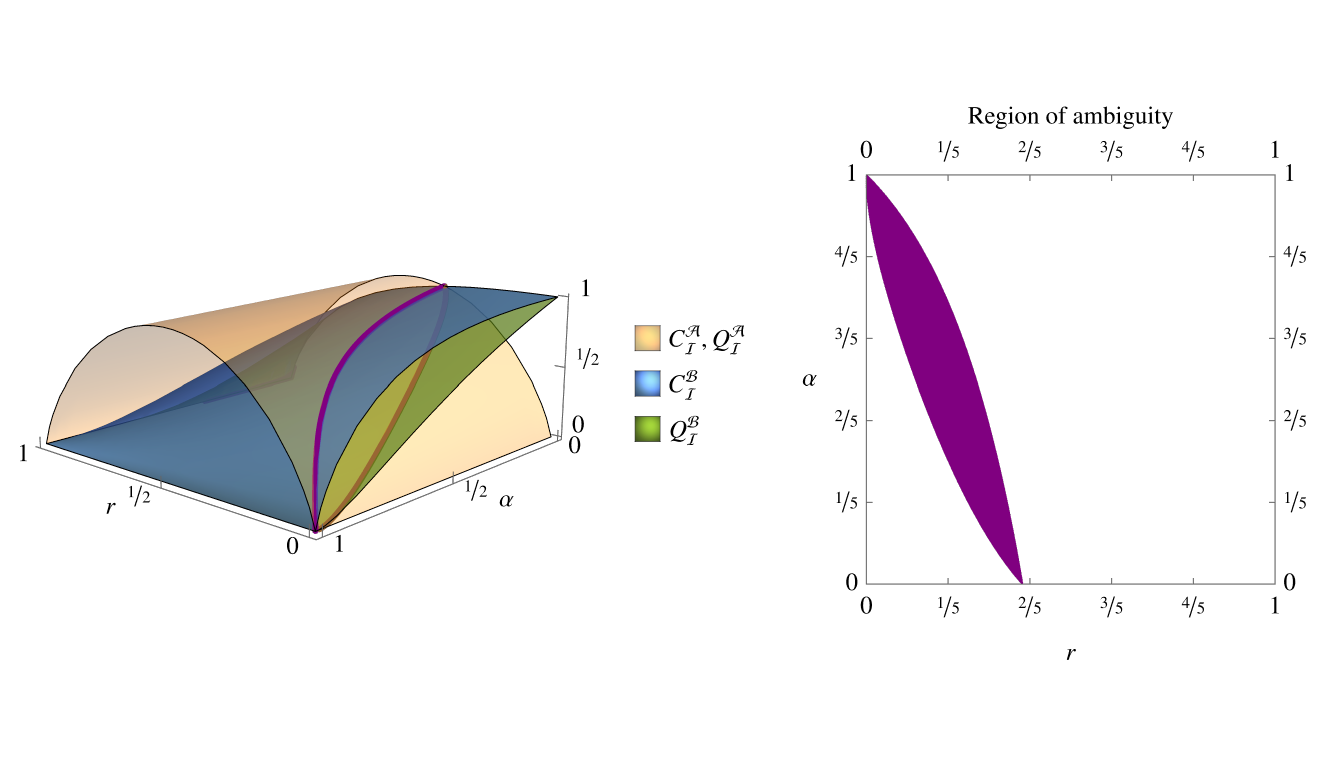}
	\caption{The classical and quantum complexities of the agents. In orange are both the classical and quantum complexity of Alice, $C_\I^\A, Q_\I^\A $. In blue is the classical complexity of Bob, $C_\I^\B $, while in green is his quantum complexity, $Q_\I^\B$. }
	\label{Fig:2agents}
\end{figure*}
\begin{figure*}[!t]
	\centering
	\begin{minipage}{0.45\textwidth}
		\centering
		\includegraphics[width=\linewidth]{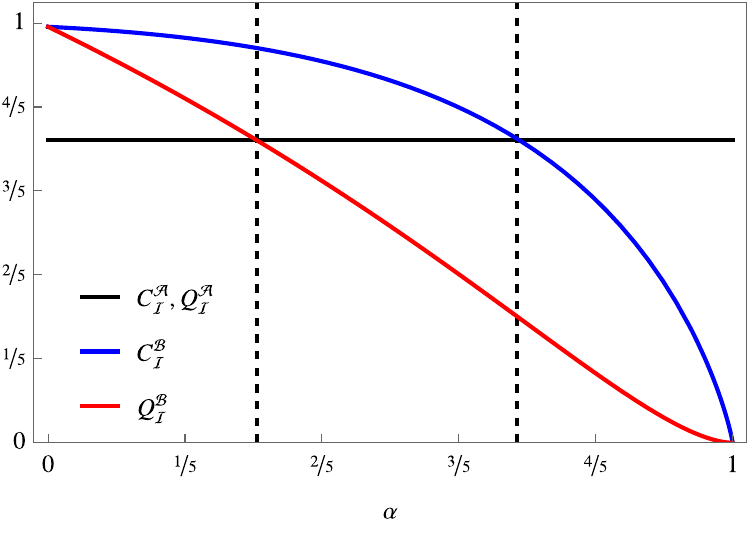} 
		\caption{Classical and quantum complexities of the agents for a biased coin input with bias $r=\nicefrac{1}{5}$. In black are the classical and quantum complexities of Alice, $C_\I^\A, Q_\I^\A $. In blue is the classical complexity, $C_\I^\B$, of Bob with varying parameter $\alpha$ while in red is his quantum complexity, $Q_\I^\B$. In the region between the dashed lines ambiguity of complexity is exhibited.}
		\label{Fig:2agents fixed r}
	\end{minipage}\hfill
	\begin{minipage}{0.45\textwidth}
		\centering
		\includegraphics[width=\linewidth]{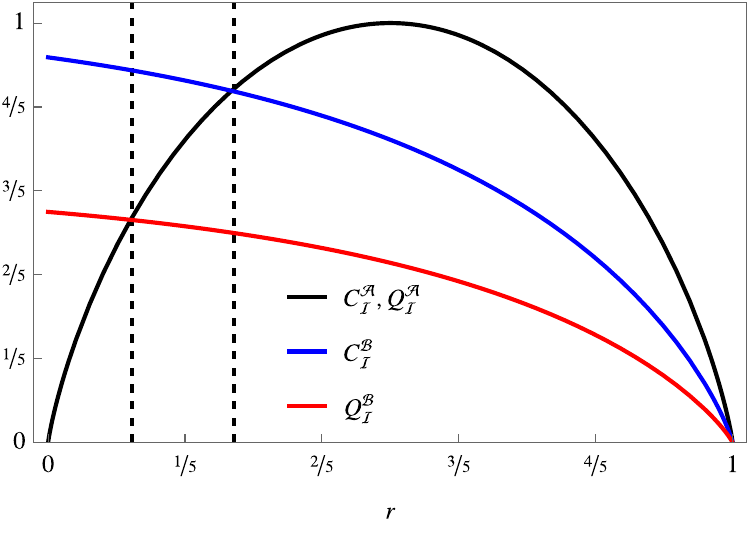}
		\caption{Classical and quantum complexities of the agents driven by a biased coin input of varying bias $r$. In black are the classical and quantum complexities of Alice, $C_\I^\A, Q_\I^\A $. In blue is the classical complexity, $C_\I^\B $, of Bob with fixed parameter $\alpha=\nicefrac{1}{2}$ while in red is his quantum complexity, $Q_\I^\B$. In the region between the dashed lines ambiguity of complexity is exhibited.}
		\label{Fig:2agents fixed alpha}
	\end{minipage}
\end{figure*}

 We now turn to the derivation of the quantum complexities. From the above optimal classical description of the agents we can construct the optimal quantum models \cite{thompson_using_2017}, \cite{elliott_quantum_2022}.  First, it is easy to show that an optimal quantum model of Alice will have the same complexity as the classical one. This follows from the fact that the fidelity between the conditional futures of the two causal states is equal to zero, forcing any quantum model to have orthogonal causal states as well (see Appendix \ref{app: 2 agents 1 stimulus}). However, this will not be the case in general for Bob. In both cases, to prove the optimality of the quantum model we show that it saturates a \emph{maximum fidelity constraint} \cite{elliott_quantum_2022}. Optimality then follows from the fact that for two states, the von Neumann entropy is a decreasing function of the overlap between the states.
Concretely, for Alice, the fidelity constraint is equal to 0 which means that the quantum states have to be orthogonal. Thus, a quantum model cannot lead to a memory reduction over the classical model and $Q_\I^\A=C_\I^\A$. This is true whenever the channel excess entropy equals the classical complexity for a certain input, and it can be shown that this always holds for processes with deterministic transitions between states. In particular, as any delay channel of $n$ time steps is deterministic, its classical and quantum complexities coincide for \emph{any} input.

For Bob we find that the maximum fidelity constraint gives the value $F_{12}= \sqrt{\alpha}$ (see Appendix \ref{app: 2 agents 1 stimulus}).
At the same time, the quantum causal states from Eqs.\@ \eqref{eq:quantum encoding 1}-\eqref{eq:quantum encoding 2}, can be encoded in a single qubit as:
\begin{align}
	&\ket{\sigma_1} =  \sqrt{\alpha} \ket{0} +\sqrt{1-\alpha}\ket{1}  \notag \\
	&\ket{\sigma_2} =\ket{0} \,.
\end{align} 
It is straightforward to verify that they obey $\bk{\sigma_1}{\sigma_2} =\sqrt{\alpha}$, thus saturating the fidelity constraint and minimising the von Neumann entropy: the quantum encoding of the quantum model is optimal.

Having obtained an optimal classical to quantum state encoding, we can now derive the quantum complexity.  In the case of Bob, the quantum complexity differs from the classical one and is explicitly found to be
\begin{align}
    Q_\I^\B = h\left(\frac{c-\sqrt{c+d}}{2c}\right)\,,
\end{align}
where $c=2+r-\alpha r$ and $d=-4r(2-4\alpha+2\alpha^2)$. Note that for Bob the classical and quantum complexities depend on both the bias of the coin $r$, as well as the parameter $\alpha$. 

We can now show that there are regions of the parameters where an inconsistency in the ordering of the complexities occurs.
This ambiguous ordering of the classical and quantum complexities of Alice and Bob is shown in Fig.\@ \ref{Fig:2agents}. In orange, we have plotted the coinciding classical and quantum complexities of Alice, while with blue and green the classical and quantum complexities of Bob, respectively. It is clear that there are values of the bias of the input $r$ and the parameter $\alpha$ of Bob, for which an inconsistent ordering occurs. The region of ambiguity of complexity is shown in purple in Fig.\@ \ref{Fig:2agents}.
In Figs.\@ \ref{Fig:2agents fixed r} and \ref{Fig:2agents fixed alpha}, we further demonstrate two specific cases where ambiguity arises. Specifically, in Fig.\@ \ref{Fig:2agents fixed r} we fix the value of the bias of the input to $r=\nicefrac{1}{5}$ and observe that any detector $\B$ with noise parameter values $\alpha$ between $0.3$ and $0.68$ exhibit ambiguity in the ordering of the complexities. Similarly, in Fig.\@ \ref{Fig:2agents fixed r} we consider a specific detector with noise parameter $\alpha = \nicefrac{1}{2}$ and vary the probability $r$ that controls whether a particle is generated or not as an input. We see that conflicting ordering between classical and quantum complexities occurs for value of the bias $r$ between $~0.12$ and $0.26$.

\begin{figure}[!t]
	\centering
	\includegraphics[width=0.9\linewidth,trim={0.1cm 0.2cm 0.1cm 0.1cm},clip]{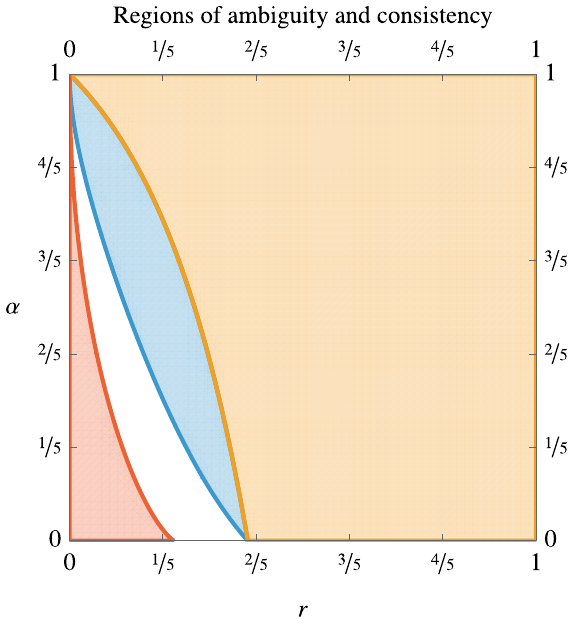}
	\caption{The regions of ambiguity and consistency as detected through the conditions of Results \ref{result: ambiguous ordering} and \ref{result: unambiguous ordering}.  Regions of ambiguity and consistency, $\R_1$ and $\R_2$, are shown in blue and orange, when the quantum complexity of $\B$, $Q_\I^\B$ is known. Region of consistency $\R_4$ is shown in red, when the quantum complexity of $\A$, $Q_\I^\A$, is known instead.}
	\label{Fig:regions of ambiguity and consistency through excess entropies}
\end{figure}

We note that the regions of ambiguity and consistency could have been predicted without having to derive both of the quantum complexities, directly by applying Results \ref{result: ambiguous ordering} and \ref{result: unambiguous ordering}. For instance, consider the case where we could only derive the optimal quantum complexity of $\B$ but not of $\A$. As the excess entropy of the input is zero, i.e. $\ee^\I = 0$, we have from Proposition \ref{prop: decomposition of E for IOP} that $\ee^\A_\I= \ee^\J-\ee^\I = \ee^\J$. As the input-output process $\A$ is deterministic, the excess entropy equals the classical (and quantum) complexity for any input. That is, $\ee^{A}_{\I}=C^{A}_{\I}=Q^{A}_{\I}=h(r)$. 
Using Result \ref{result: ambiguous ordering}, we define the region of ambiguity, $\mathcal{R}_1$, the region in the parameter space $(\alpha,r)$ 
\begin{align}
    \mathcal{R}_1 : \{C_\I^\B > C_\I^\A \text{ and } \ee^{\A}_\I \geq Q_\I^\B \} \,. \label{eq: region 2}
\end{align}
Similarly, from Result \ref{result: unambiguous ordering} we can define the region of consistency of the complexities, $\mathcal{R}_2$, as
\begin{align}
    \mathcal{R}_2 : \{C_\I^\A > C_\I^\B \text{ and } \ee^{\A}_\I \geq Q_\I^\B \} \,. \label{eq: region 1}
\end{align}
The two regions $\mathcal{R}_1$ and $\mathcal{R}_2$ are shown in blue and orange, respectively, in Fig.\@ \ref{Fig:regions of ambiguity and consistency through excess entropies}.

Were we able to evaluate the quantum complexity of $\A$ but not that of $\B$, Results \ref{result: ambiguous ordering}
 and \ref{result: unambiguous ordering} would imply instead two regions $\mathcal{R}_3$ and $\mathcal{R}_4$ of ambiguity and consistency, respectively, defined as
 \begin{align}
    \mathcal{R}_3 : \{C_\I^\A > C_\I^\B \text{ and } \ee^{\B}_\I \geq Q_\I^\A \}
    \label{eq: region 3}
\end{align}
and
\begin{align}
    \mathcal{R}_4 : \{C_\I^\B > C_\I^\A \text{ and } \ee^{\B}_\I \geq Q_\I^\A \} \,. \label{eq: region 4}
\end{align}
The excess entropies are given in Appendix \ref{app: excess entropies scenario A}. The region $\mathcal{R}_4$ is shown in red in Fig.\@ \ref{Fig:regions of ambiguity and consistency through excess entropies}, while the conditions of $\mathcal{R}_3$ are not satisfied for any values of $\alpha$ an $r$. The blue, orange and red regions in Fig.\@ \ref{Fig:regions of ambiguity and consistency through excess entropies} cover the largest area in the parameter space, demonstrating that the sufficient conditions of our Results \ref{result: ambiguous ordering} and \ref{result: unambiguous ordering} are able to successfully detect most of the paramater space. In the white region of Fig.\@ \ref{Fig:regions of ambiguity and consistency through excess entropies},  Results \ref{result: ambiguous ordering} and \ref{result: unambiguous ordering} are agnostic about the ordering of the complexities.

Having demonstrated this ambiguous ordering in the complexities, it is natural to ask how common this phenomenon is. We have the following result, which shows that this phenomenon is not isolated.

\begin{result}
\label{res: continuous with C=Q in the gap}
Given an input stochastic process $\I$ and an input-output process $\A$ with non-equal optimal classical and quantum complexities when driven by $\I$, i.e. $Q^{\A}_{\I}<C^{\A}_\I$, there exists a family of input-output processes $\B_{\vec{q}}$ with equal classical and quantum complexities for any value of the multi-parameter vector $\vec{q}$ when driven by $\I$, i.e. $C^{\B_{\vec{q}}}_\I=Q^{\B_{\vec{q}}}_\I$, for any value in $[Q^{\A}_{\I},C^{\A}_\I]$. Precisely, $C^{\B_{\vec{q}}}_\I=Q^{\B_{\vec{q}}}_\I=C^{\A}_\I-Q^{\A}_{\I}+\varepsilon$, for all $\varepsilon \in [Q^{\A}_{\I},C^{\A}_\I]$ and for some choice of $\vec{q}$.
\end{result}

That is, given an agent who implements an adaptive strategy of which the optimal quantum implementation has lower complexity than the optimal classical one, there exist continuous-labelled adaptive strategies that another agent can implement that have equal classical and quantum complexities, and which have value lying in the range between the classical and quantum complexity of the first agent. Thus, for each such pair of adaptive strategies, an ambiguous ordering of classical and quantum complexities will occur. 
Given that in the case of stochastic processes, those that have a gap between their classical and quantum complexities seem to be the norm instead of the exception \cite{guQuantumMechanicsCan2012,ellison_information_2011} and anticipating the same to hold for input-output processes, we see that this phenomenon is very common.
In Appendix\@ \ref{app: input-output processes that fill gap}, we prove Res.\@ \ref{res: continuous with C=Q in the gap} by constructing such a family of input-output process and show that
(i) their classical and quantum complexities are the same and (ii) by adjusting the free parameters, their complexities can attain any value in $\mathbb{R}$.

\subsection{A single agent reacting to two different stimuli}

The same phenomenon persists even for a single agent implementing a strategy reacting to different input stimuli. That is, a \emph{single} agent executing the \emph{same} strategy may not be able to consistently assign an ordering of the complexities for two different input stimuli, when implementing the strategy classically or quantum mechanically. 

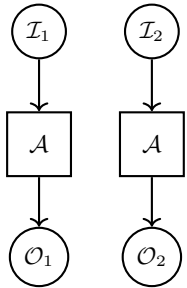
\begin{figure}[!b]
	\centering
	\begin{tikzpicture}[node distance=1.5cm, thick,square/.style={regular polygon,regular polygon sides=4,minimum size=1.2cm}, main/.style={draw, circle,minimum size=0.7cm}]
		\node[main] (1) {$\I_1$}; 
		\node [main] [right of=1] (2) {$\I_2$};
		\node[square, draw] [below of=1] (3) {$\A$}; 
		\node[square, draw] [right of=3] (4) {$\A$};
		\node[circle, draw]  [below of=3] (5) {$\O_1$}; 
		\node[circle, draw] [right of=5] (6) {$\O_2$};
		\draw[->] (1) -- (3);
		\draw[->] (2) -- (4);
		\draw[->] (3) --(5);
		\draw[->] (4) --(6);
	\end{tikzpicture} 
	\caption{A single agent reacting to two different stimuli.}
	\label{Fig:one agent}	
\end{figure}

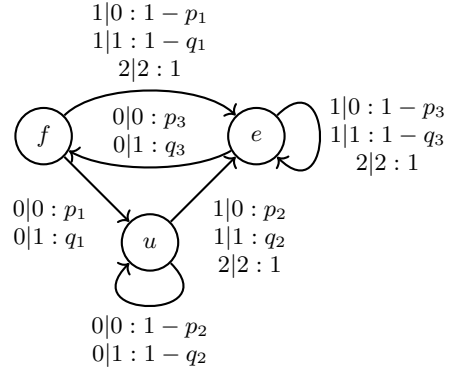
\begin{figure}[!t]
	\centering
	\scalebox{1}{
		\begin{tikzpicture}[node distance={20mm}, thick, main/.style = {draw, circle,minimum size=0.75cm}] 
			\node[main] (1) {$f$}; 
			\node[main] (3) [below right of=1] {$u$}; 
			\node[main] (2) [above right of=3] {$e$}; 
			
			\draw[->] (1) -- node[align=center,below left] {$0|0:p_1$ \\ $0|1:q_1$ }  (3); 
			\draw[->] (1) to [out=45,in=135,looseness=0.7] node[align=center, midway, above] {$1|0:1-p_1$ \\ $1|1:1-q_1$  \\ $2|2:1$}  (2); 
			
			\draw[->] (2) to [out=210,in=-30,looseness=0.7] node[align=center,above] {$0|0:p_3$ \\ $0|1:q_3$ }  (1); 
			\draw[->] (2) to [out=45,in=-45,looseness=5] node[align=center,midway,right] { $1|0:1-p_3$ \\ $1|1:1-q_3$ \\ $2|2:1$} (2);
			
			\draw[->] (3) -- node[align=center,below right] {$1|0:p_2$ \\ $1|1:q_2$ \\ $2|2:1$}  (2); 
			\draw[->] (3) to [out=-45,in=-135,looseness=5] node[align=center,midway,below] {$0|0:1-p_2$ \\ $0|1:1-q_2$} (3);		
		\end{tikzpicture} 
	}
	\caption{A 6-parameter family of input-output processes that can exhibit ambiguous ordering between classical and quantum complexities.}
	\label{Fig:etr A to B}
\end{figure}

Concretely, consider a strategy $\A$ that consists of three internal states and a three symbol input and output alphabets. Their generic behaviour can be briefly summarised as follows: (i) at each state whether an input of 0 or 1 is received a noisy output of 0 or 1 is output with probabilities that depend on both the input and present states, while if a 2 is received always a 2 is output and always a transition to state 2 follows, (ii) states 2 and 3 self-transition on some inputs while state 1 always transitions to states 2 or 3. 
The \etr{} of the input-output process is shown in Fig.\@ \ref{Fig:etr A to B}.
We note that the process is of Markov order 2 as two past outputs uniquely determine the current internal state of the input-output process.

A physical interpretation of the above setting is a scenario of an investor investing their money based on market sentiment. The inputs the investor receives represent the market sentiment they observe; specifically, the input symbols 0,1,2 represent `bad news', `mixed news', and `good news', respectively. The outputs of the investor correspond to the amounts they are willing to invest; specifically, symbols 0,1,2 represent no investment, moderate investment, and big investment, respectively. Finally, the internal states in the graph in Fig.\@ \ref{Fig:etr A to B} represent the current \emph{mood} of the investor, with labels $e$, $u$ and $f$ denoting the investor being enthusiastic about the prospect of investing, unlikely to invest, or frustrated.

Consider the case where the investor is reacting in two scenarios where the market sentiment is IID: the news in the two scenarios occur with probabilities $\I_1^{(x)}=\{\nicefrac{2}{10},\nicefrac{1}{10},\nicefrac{7}{10}\}$ and $\I_2^{(x)}=\{\nicefrac{1}{10},\nicefrac{7}{10},\nicefrac{2}{10}\}$ for the results $x=0,1,2$, respectively. Note that the two input processes are the same up to a cyclic relabelling of the alphabet. Regardless of being related by a simple relabelling, they suffice to generate inconsistency between the case the investor is implementing strategy $\A$ classically versus quantum mechanically. 
We further restrict the parameters of strategy $\A$ to the values $p_1=p_2=0, p_3=\nicefrac{4}{7}, q_2=\nicefrac{3}{5},q_3=\nicefrac{1}{100}$ while leaving $q_1$ as a free parameter.
We now derive the optimal quantum model and subsequently explicitly evaluate the classical and quantum complexities for the two inputs and finally show that there is a range of values of $q_1$ such that the difference of the classical complexities is positive, $C_{\I_1}^\A - C_{\I_2}^\A >0 $, while the difference of quantum complexities is negative, $Q_{\I_1}^\A - Q_{\I_2}^\A<0$.

To derive the optimal quantum model, we first evaluate the maximum fidelity constraint for each pair of causal states, that is, the maximum allowed value of the overlap that the quantum states of a quantum model can attain while faithfully reproducing the behaviour of the classical model. Specifically, we find the values (see Appendix\@ \ref{app: 1 agent 2 stimuli})
\begin{align}
	F_{fu}&=F_{eu}=0 \notag \\
	F_{fe} &= \min\left\{\sqrt{\frac{4}{7}}, \frac{\sqrt{99(1-q_1)}}{10}\right\} .
\end{align}
From the first two it follows that one of the states must be orthogonal to the other two in the quantum model as well. In other words, the quantum causal states $\ket{\sigma_i}=\{\ket{f}, \ket{e}, \ket{u}\}$ must obey $\braket{f}{u}=\braket{e}{u}=0$, while $\braket{f}{e}=c$ with $\abs{c}\in [0,F_fe]$. That is, we need to construct a triple of states obeying the above constraints with a value of $c$ that minimises the entropy.

Assume that the quantum agent is driven by an input that leads to a stationary distribution over the causal states of the \etr{} with probabilities $\pi=\{\pi_f,\pi_e,\pi_u\}$. Then, the quantum complexity can be calculated through the eigenvalues of the Gram matrix, which shares non-zero eigenvalues with the state $\rho = \sum_{x\in \{f,e,u\}} \pi_x \ketbra{x}{x}$ \cite{jozsa_distinguishability_2000}.  The Gram matrix in this case is 
\begin{align}
	G = \begin{pmatrix}
		\pi_f & \sqrt{\pi_f \pi_e} c & 0 \\
		 \sqrt{\pi_f \pi_e} c^* & \pi_e & 0 \\
		 0 & 0 & \pi_u  
	\end{pmatrix} \,,
\end{align}
and with $\pi_u=1-\pi_f+\pi_e$ its eigenvalues are found to be 
\begin{align}
	e_i =\{&1-\pi_f-\pi_e \,,  \notag \\
    &\frac{1}{2} \left( \pi_f+\pi_e \pm \sqrt{(\pi_1-\pi_e)^2+4\abs{c}^2 \pi_f \pi_e} \right)\}\,.
\end{align} 
As the von Neumann entropy has a strictly negative derivative with respect to $\abs{c}$, it is a decreasing function of $\abs{c}$. Thus the quantum complexity is minimised for the maximum allowed overlap, $\abs{c}=F_{fe}$. 
Next, one needs to construct a set of quantum states that achieve these overlaps. It is straightforward to check that such a triple is
\begin{align}
	\ket{f}&=  \ket{1}, \notag \\
	\ket{e}&= F_{fe}\ket{1}+\sqrt{1-F_{fe}^2}\ket{2} \notag \\
	\ket{u}&= \ket{3},
\end{align}
where $\{\ket{j}\}_{j=1,\ldots,3}$ denotes an orthonormal basis. 
\begin{figure}[!t]
	\centering
	\includegraphics[width=1\linewidth]{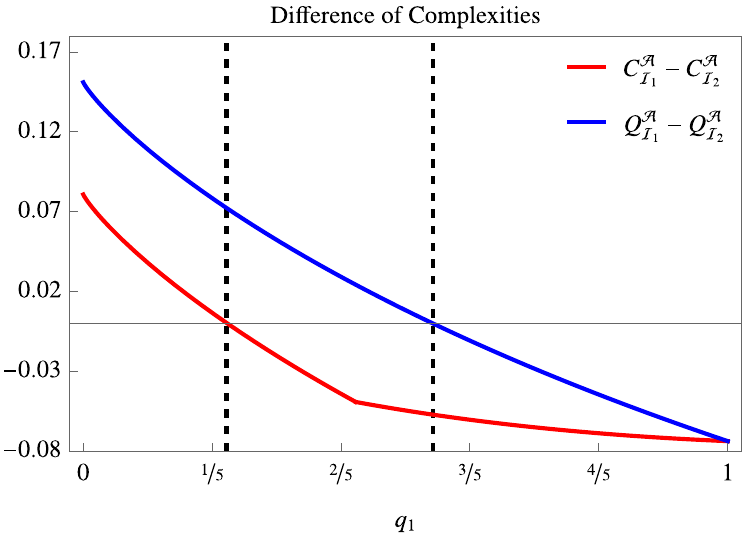}
	\caption{The difference of classical and quantum complexities of the adaptive strategies of two agents reacting to different stimuli as a function of the free parameter $q_1$. In blue we plot the difference of the classical complexities, $C_{\I_1}^\A - C_{\I_2}^\A $, while in red we show the difference of quantum complexities in the two cases, $Q_{\I_1}^\A - Q_{\I_2}^\A$. The vertical dashed lines represent the points that the graphs intersect with the horizontal line at zero, thus changing sign. }
	\label{Fig:1agent}
\end{figure} 
In Fig.\@ \ref{Fig:1agent} we plot the difference of the classical and quantum complexities as a function of the parameter $q_1$ of the input-output process. We find that there is a class of such processes, i.e., a region of values of $q_1$, with inconsistent ordering. Specifically, if the free parameter $q_1$ in the input-output process takes value in $0.22\leq q_1\leq 0.54$, the two graphs have opposite signs, and as a result inconsistent ordering of the classical and quantum complexities occurs.

\begin{result}
    There exists an input-output process $\A$ and two input processes $\I_1$ and $\I_2$ such that the difference of the classical complexities of $\A$ when driven by $\I_1$, $C_{\I_1}^\A$, and when driven by $\I_2$, $C_{\I_2}^\A$, is positive, $C_{\I_1}^\A - C_{\I_2}^\A >0$,  while the difference of the respective quantum complexities $Q_{\I_1}^\A$ and $Q_{\I_2}^\A$ is negative, $Q_{\I_1}^\A - Q_{\I_2}^\A <0$.
\end{result}

It is natural to ask whether given an input-output process that can exhibit different classical and quantum complexities for some input, there are always input processes that can lead to an inconsistent ordering of the complexities. We only have the following partial answer.
\begin{observation}
    There exist input-output processes that can have non-equal classical and quantum complexities, which however cannot lead to ambiguity in the ordering of the complexities for any pair of IID input processes. 
\end{observation}
We demonstrate this observation with a specific example in Appendix\@ \ref{app: input output processes with no ambiguity IID}.

\subsection{An Agent and Their Operational Inverse}

Given an agent that reacts to inputs from an input process $\I$ and executing an input-output process $\A$, their output actions also define a stochastic process $\O$. Thus, in effect the input-output process $\A$ that the agent is implementing maps the input process $\I$ to an output process $\O$. Often, an agent that implements an \emph{inverse} transformation which maps $\O$ back to $\I$ is useful.  Here we consider such an agent that implements an \emph{inverse} input-output process, more formally defined as follows: 
\begin{defn}
    An \emph{inverse} of an input-output process $\A$ for a given input $\I$ which is mapped to an output process $\O$ though $\A$, is an input-output process $\A^{-1}$ such that when driven by the stochastic process $\O$ it maps to the original stochastic process $\I$. 
    \label{def: inverse IO}
\end{defn}

Although an input process $\I$ and an input-output $\A$, do not uniquely specify an inverse process in general, there is a way to infer such inverse channels.
This can be done by considering the implied stochastic process over the joint inputs and outputs. From the joint and output stochastic processes one can infer an input-output process that implements the inverse map that takes outputs to inputs by appropriately conditionalising the joint process over outputs. Note that, in general, one cannot infer a unique channel this way, and thus a class of different input-output processes will be obtained. Moreover, in some cases the map from outputs back to inputs could be completely trivial, in the sense that it will be equivalent to ignoring all inputs and just outputting the target output process. Regardless, in this case, too, we show that ambiguous ordering of the complexities between the maps can exist. 
We remark that here we will show that this ambiguity can exist between an agent implementing an input-output process for a given input stimulus and an agent implementing the inverse actions, but a closely related scenario was considered in Ref.\@ \cite{kechrimparis_causal_2023}. There, given a pair of stochastic processes, it was shown that this phenomenon persists for the \emph{memory minimal} channels between the processes, which demonstrated an inconsistent \emph{causal asymmetry}. Thus, in the scenario considered here we start with an input stochastic process and an input-output process as given, while in Ref.\@ \cite{kechrimparis_causal_2023} the starting point was a pair of stochastic processes. In Ref.\@ \cite{kechrimparis_causal_2023} minimisation of the classical and quantum complexities over all possible forward and inverse input-output processes was considered, while here we have an asymmetric scenario as one of the input-output processes is fixed and, in addition, minimisation over all inverse input-output processes is not demanded.

Consider an input process $\I$ as well as an input-output process $\A$. If $M^{(x)}$ denote the transition matrices of $\I$ and $T^{(y|x)}$ are the transition matrices describing the input-output process $\A$, then we can derive the transition matrices of the process over joint inputs and outputs, $\J$, through
\begin{align}
    J^{(x,y)}=T^{(y|x)}\otimes M^{(x)} \,. \label{eq: joint process from input and i/o}
\end{align}
By marginalising over inputs, we can obtain a presentation of the outputs of the input-output process, $\O$,
\begin{align}
    N^{(y)}=\sum_x T^{(y|x)}\otimes A^{(x)} \,, \label{eq: output process from marginalisation}
\end{align}
where $N^{(y)}$ denote the transition matrices of $\O$.
Note that in both cases some states may be transients, i.e. they have zero long-term probability of occurrence, and thus can be removed. Moreover, the resulting presentation may not be minimal, so one would also need to check for equivalence of the states \cite{james_many_2014}. 
\begin{figure*}[!t]
    \centering

    \begin{minipage}{0.45\textwidth}
        \centering
        \scalebox{1}{
	\begin{tikzpicture}[node distance={15mm}, thick, main/.style = {draw, circle,minimum size=0.7cm}] 
		\node[main] (1) {$r_0$}; 
		\node[main] (2) [ right of=1] {$r_1$}; 
		
		\draw[->] (1) to [out=45,in=135,looseness=1] node[align=center,midway,above] {$1:\nicefrac{1}{2}$\\$2:\nicefrac{1}{2}$}  (2); 
		\draw[->] (2) to [out=-135,in=-45,looseness=1] node[align=center,midway,below] {$0:\nicefrac{1}{2}$}  (1); 
		\draw[->] (2) to [out=45,in=-45,looseness=5] node[align=center,midway,right] {$1:\nicefrac{1}{2}$ } (2);
	\end{tikzpicture} 
    }
	\caption{The input stimulus described by the process $\I$ .}
	\label{Fig:input stimulus A}
\end{minipage} 
\begin{minipage}{0.45\textwidth}
        \centering
	\centering
	\scalebox{0.95}{
		\begin{tikzpicture}[node distance={28mm}, thick, main/.style = {draw, circle,minimum size=0.7cm}] 
			\node[main] (1) {$\sigma_0$}; 
			\node[main] (3) [below right of=1] {$\sigma_1$}; 
			\node[main] (2) [above right of=3] {$\sigma_2$}; 
			
			\draw[->] (3) -- node[align=center, right, pos=0.7] {$0|0:1$}  (1); 
			\draw[->] (1) to [out=45,in=135,looseness=0.7] node[align=center, midway, above] {$2|1:1-p$ \\ $2|2:1$}  (2); 
			\draw[->] (1) to [out=-90,in=180,looseness=0.7] node[align=center, midway, below left] {$1|1:p$ }  (3); 
			
			\draw[->] (2) to [out=180,in=-0,looseness=0] node[align=center,above] {$0|0:1$ }  (1); 
			\draw[->] (2) -- node[pos=0.3,align=center,left] {$1|1:r$ }  (3); 
			\draw[->] (3) to [in=-90,out=0,looseness=0.7] node[align=center, below right] {$2|1:1-q$ \\ $2|2:1$ }  (2); 
			
			\draw[->] (1) to [out=225,in=135,looseness=5] node[align=center,midway,left] {$0|0:1$} (1);
			\draw[->] (2) to [out=45,in=-45,looseness=5] node[align=center,midway,right] { $2|1:1-r$ \\ $2|2:1$} (2);
			\draw[->] (3) to [out=-45,in=-135,looseness=5] node[align=center,midway,below] {$1|1:q$} (3);
		\end{tikzpicture} 
	}	
	\caption{The \etr{} of an input-output process $\A$ with three states.}
    \label{Fig:Markovian agent}
    \end{minipage}
    \begin{minipage}{0.45\textwidth}
	\scalebox{1}{
		\begin{tikzpicture}[node distance={28mm}, thick, main/.style = {draw, circle,minimum size=0.7cm}] 
			\node[main] (1) {$s_0$}; 
			\node[main] (3) [below right of=1] {$s_1$}; 
			\node[main] (2) [above right of=3] {$s_2$}; 
			
			\draw[->] (3) -- node[align=center, right, pos=0.4] {$(0,0):\frac{1}{2}$}  (1); 
			\draw[->] (1) to [out=45,in=135,looseness=0.7] node[align=center, midway, above] {$(1,2):\frac{1-p}{2}$ \\ $(2,2):\frac{1}{2}$}  (2); 
			\draw[->] (1) to [out=-90,in=180,looseness=0.7] node[align=center, midway, below left] {$(1,1):\frac{p}{2}$ }  (3); 
			
			\draw[->] (2) to [out=180,in=-0,looseness=0] node[align=center,above] {$(0,0):\frac{1}{2}$ }  (1); 
			\draw[->] (2) -- node[pos=0.2,align=center,left] {$(1,1):\frac{r}{2}$ }  (3); 
			\draw[->] (3) to [in=-90,out=0,looseness=0.7] node[align=center, below right] {$(1,2):\frac{1-q}{2}$ }  (2); 
			
			\draw[->] (2) to [out=45,in=-45,looseness=5] node[align=center,midway,right] { $(1,2):\frac{1-r}{2}$} (2);
			\draw[->] (3) to [out=-45,in=-135,looseness=5] node[align=center,midway,below] {$(1,1):\frac{q}{2}$} (3);	
		\end{tikzpicture} 
	}
	\caption{The \ema{} of the joint process $\J$ over inputs and outputs.}
    \label{Fig:joint process}
    \end{minipage}\quad \quad
    \begin{minipage}{0.45\textwidth}
	\scalebox{1}{
		\begin{tikzpicture}[node distance={28mm}, thick, main/.style = {draw, circle,minimum size=0.7cm}] 
			\node[main] (1) {$\chi_0$}; 
			\node[main] (3) [below right of=1] {$\chi_1$}; 
			\node[main] (2) [above right of=3] {$\chi_2$}; 
			
			\draw[->] (3) -- node[align=center, right, pos=0.6] {$0:\frac{1}{2}$}  (1); 
			\draw[->] (1) to [out=45,in=135,looseness=0.7] node[align=center, midway, above] {$2:1-\frac{p}{2}$}  (2); 
			\draw[->] (1) to [out=-90,in=180,looseness=0.7] node[align=center, midway, below left] {$1:\frac{p}{2}$ }  (3); 
			
			\draw[->] (2) to [out=180,in=-0,looseness=0] node[align=center,above] {$0:\frac{1}{2}$ }  (1); 
			\draw[->] (2) -- node[pos=0.4,align=center,left] {$1:\frac{r}{2}$ }  (3); 
			\draw[->] (3) to [in=-90,out=0,looseness=0.7] node[align=center, below right] {$2:\frac{1-q}{2}$ }  (2); 
			
			\draw[->] (2) to [out=45,in=-45,looseness=5] node[align=center,midway,right] { $2:\frac{1-r}{2}$} (2);
			\draw[->] (3) to [out=-45,in=-135,looseness=5] node[align=center,midway,below] {$1:\frac{q}{2}$} (3);
		\end{tikzpicture} 
	}
	\caption{The \ema{} $\O$ of the process describing the outputs of the input-output process $\A$ when driven by $\I$.}
    \label{Fig:output process}	
    \end{minipage}
\end{figure*}

Given the output $\O$, we can conditionalise $\J$ to get a presentation of the inverse input-output process that upon receiving inputs from $\O$, it outputs $\I$.
This can be achieved by looking at the transitions between states of $\J$ and using the probabilities from the corresponding transitions from $\O$ to obtain the conditional probabilities of the postulated inverse input-output process, $\A^{-1}$. Specifically, we have the following inversion algorithm.
\begin{alg}
    Given an input-output process $\A$ driven by some input process $\I$, and producing outputs that define an output process $\O$, inverse input-output processes $\A^{-1}$ that map $\O$ back to $\I$ can be derived with the following steps
    \begin{enumerate}
        \item For a pair of states $s_i$ and $s_j$ of $\J$, identify the corresponding states $\chi_k, \chi_l$ in the output process, $\O$.
        \item Define the postulated inverse input-output states with labels $(s_i, \chi_k)$ and $(s_j, \chi_l)$.
        \item For each transition between states $s_i, s_j$ of $\J$, divide the probability in all emissions $(x,y)$ with the corresponding probability of emission $y$ between $\chi_k$ and $\chi_l$ of $\O$ and obtain the probabilities of conditional emissions $x|y$ between $(s_i, \chi_k)$ and $(s_j, \chi_l)$ of $\A^{-1}$.
        \item Repeat until all pairs of states of $\J$ have been considered.
    \end{enumerate}
\end{alg}
Note that this algorithm will give an input-output process that maps $\O$ to $\I$ but there may be undefined transitions, which can are essentially free parameters. This reflects the fact that there is no unique channel between two stochastic processes. In addition, it is possible that some of the states of the inverse input-output process are equivalent. Thus, to obtain a minimal presentation, that is, the \etr{} of an inverse input-output process, the states have to be checked for equivalence \cite{james_many_2014}. 

We have thus shown that
\begin{result}
    Given an input-output process $\A$ which maps an input stochastic process $\I$ to a stochastic process $\O$, a family of inverse input-output processes $\A^{-1}$ that map $\O$ back to $\I$ can be found through Algorithm 1. 
\end{result}

We demonstrate this result with an explicit example.
Consider the input stimulus $\I$ whose \ema{} is shown in Fig.\@ \ref{Fig:input stimulus A} and the input-output process $\A$ shown in Fig.\@ \ref{Fig:Markovian agent}. Note that the input is Markovian since each emitted symbol identifies a unique state of the \ema{}. Similarly, the input-output process is Markovian on output symbols.
The joint and output process can be found through Eqs.\@ \eqref{eq: joint process from input and i/o} and \eqref{eq: output process from marginalisation}. To derive their \emas{}, states need to be checked for equivalence \cite{james_many_2014}.
In the particular case that we are examining, all the resulting states are inequivalent. The \emas{} of the resulting processes $\J$ and $\O$ are shown in Figs.\@ \ref{Fig:joint process} and \ref{Fig:output process}.

\begin{figure}[!b]
	\centering
	\begin{tikzpicture}[node distance={20mm}, thick, main/.style = {draw, circle,minimum size=0.7cm}] 
	\node[main] (1) {$\phi_1$}; 
	\node[main] (2) [ right of=1] {$\phi_0$}; 
			
	\draw[->] (1) to [out=45,in=135,looseness=1] node[align=center,midway,above] {$0|0:1$ } (2); 
	\draw[->] (2) to [out=-135,in=-45,looseness=1] node[align=center,midway,below] {$1|1:1$ \\ $1|2:\frac{1-p}{2-p}$ \\ $2|2:\frac{1}{2-p}$ }  (1); 
	\draw[->] (1) to [out=225,in=135,looseness=5] node[align=center,midway,left] {$1|1:1$ \\ $1|2:1$  } (1);
    \draw[->] (2) to [out=45,in=-45,looseness=5] node[align=center,midway,right] {$0|0:1$} (2);	
	\end{tikzpicture} 
	\caption{The \etr{} of the implied inverse input-output process $\A^{-1}$ that maps $\O$ back to $\I$, with $\phi_0=(s_0,\chi_0)$ and $\phi_1=(s_1,\chi_1)$.}
	\label{Fig:implied iverse agent}
\end{figure}
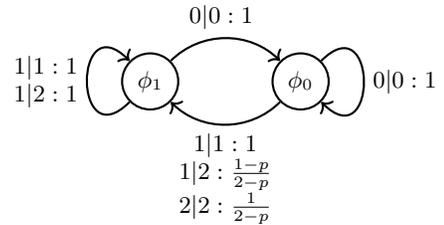

We now apply Algorithm 1 to find an inverse input-output process of $\A$.
We first note that in the example we are considering, careful inspection reveals that a state $s_i$ of $\J$ corresponds to a state $\chi_i$ of $\O$.
First, we consider the transition from state $s_0$ to $s_2$ of $\J$, which implies the two conditional transitions of $1|2:\frac{1-p}{2-p}$ and $2|2: \frac{1}{2-p}$. Repeating for all states, we find
\begin{itemize}
    \item $s_0$ to $s_2$: $1|2:\frac{1-p}{2-p}$ and $2|2: \frac{1}{2-p}$ and $s_0$ to $s_1$: $1|1:1$
    \item $s_1$ to $s_1$: $1|1:1$, $s_1$ to $s_2$: $1|2:1$ and $s_1$ to $s_0$: $0|0:1$.
    \item $s_2$ to $s_2$: $1|2:1$, $s_2$ to $s_1$: $1|1:1$ and $s_2$ to $s_0$: $0|0:1$.
\end{itemize}
Noting that the \etr{} of the inverse agent is Markovian and checking for equivalence, we find that the states $(s_1,\chi_1)$ and $(s_2,\chi_2)$ are in fact equivalent and can be merged. To check for equivalence we need to check that for all inputs, all outputs words of length one are statistically indistinguishable \cite{james_many_2014,travers_exact_2011}. Finally, note that from state $s_0$ there is no implied transition over $0$ as input. This means that whichever way we add this transition, it will not affect the output. We choose to add a self transition of $s_0$ as $0|0:1$. 
The resulting \etr{} is shown in Fig. \ref{Fig:implied iverse agent}. 

When it comes to the ordering of classical and quantum complexities, we have the following result:
\begin{result}
    There exist an input stochastic process $\I$ and an input-output process $\A$ that maps $\I$ to some output stochastic process $\O$, and inverse maps $\A^{-1}$ of $\A$ in the sense of Definition \ref{def: inverse IO}, such that that the difference of the classical complexities of $\A$ when driven by $\I$, $C_{\I}^\A$, and of its inverse $\A^{-1}$ when driven by $\O$, $C_{\O}^{\A^{-1}}$, is positive, $C_{\I}^\A - C_{\O}^{\A^{-1}} >0$, while the difference of the respective quantum complexities $Q_{\I}^\A$ and $Q_{\O}^{\A^{-1}}$ is negative, $Q_{\I}^\A - Q_{\O}^{\A^{-1}} <0$.
\end{result}

To demonstrate the ambiguous ordering between the complexities, we take the following values of the free parameters of input-output process $\A$: $p=0, \, q=\nicefrac{1}{3}, \, r=\nicefrac{1}{4}$. A direct evaluation of the stationary distributions of $\A$ and $\A^{-1}$ leads to the classical complexities $C_{\I}^{\A}=1.290$ and $C_{\O}^{\A^{-1}}=0.918$, which implies that $C_{\I}^{\A}>C_{\O}^{\A^{-1}}$, that is, a classical agent executing $\A$ is more complex than a classical agent implementing its operational inverse, $\A^{-1}$. On the other hand, encoding the quantum states as in Eqs.\@ \eqref{eq:quantum encoding 1}-\eqref{eq:quantum encoding 2} and calculating the quantum complexities, we find the values $Q_{\I}^{\A}=0.336$ and $Q_{\O}^{\A^{-1}}=0.550$. Note that for the case of $\A^{-1}$, the quantum causal states saturate the maximum fidelity constraint and thus correspond to an optimal encoding. A similar statement about optimality can not be made about the encoding of $\A$, but since the optimal quantum model will have at most the same complexity as our encoding, it suffices. We conclude that $Q_{\I}^{\A}<Q_{\O}^{\A^{-1}}$, which shows that a quantum agent implementing $\A$ will have less complexity than a quantum agent implementing the inverse actions $\A^{-1}$, leading to inconsistent ordering.

\section{Complexity of strategies vs complexity of outputs}
\label{sec: outputs vs strategies}
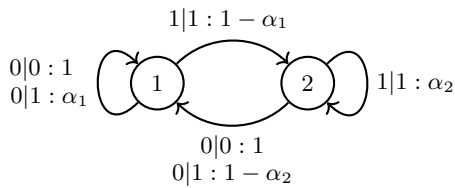
\begin{figure}[!b]
	\centering
	\begin{tikzpicture}[node distance={20mm}, thick, main/.style = {draw, circle,minimum size=0.7cm}] 
		\node[main] (1) {$1$}; 
		\node[main] (2) [ right of=1] {$2$}; 
		
		\draw[->] (1) to [out=45,in=135,looseness=1] node[align=center,midway,above] {$1|1:1-\alpha_1$ } (2); 
		
		\draw[->] (2) to [out=45,in=-45,looseness=5] node[align=center,midway,right] {$1|1:\alpha_2$ } (2); 
		\draw[->] (2) to [out=-135,in=-45,looseness=1] node[align=center,midway,below] {$0|0:1$\\$0|1:1-\alpha_2$}  (1); 
		\draw[->] (1) to [out=225,in=135,looseness=5] node[align=center,midway,left] {$0|0:1$ \\$\,\,\,\,0|1:\alpha_1$  } (1);
		
	\end{tikzpicture} 
	\caption{Input-output process that maps a biased coin to the Ising spin chain process.}
	\label{Fig:Agent 3}
\end{figure}

In the previous sections, we showed that inconsistent ordering may exist between the classical and quantum complexities of agents implementing different input-output processes. We found that the outputs of an input-output process driven by an input stochastic process, when viewed alone, define a stochastic process, the output process. It is natural to ask whether this inconsistency exhibited by the agents is always reflected on the output processes as well? In other words, are the optimal classical and quantum complexities of the output processes also inconsistent? For the examples considered in the previous section, it can be shown that the complexities of the output process follow the complexities of the input-output processes for IID inputs. However, we will show that this is not the case in general. In fact, all combinations are possible: (i) consistent input-output - consistent outputs, (ii) inconsistent input-output - inconsistent outputs, (iii) consistent input-output - inconsistent outputs, (iv) inconsistent input-output - consistent outputs. Case (i) can be trivially demonstrated by taking two input-output processes without a gap between the classical and quantum complexities, driven by appropriate inputs that lead to output processes that also do not have a gap between the classical and quantum complexities. We will demonstrate cases (ii)-(iv) for a scenario of two input-output processes driven by the same stimulus.

We consider the family of input-output processes shown in Fig.\@ \ref{Fig:Agent 3}, which map a biased coin to the Ising Spin chain process \cite{suen_classical-quantum_2017}.  The input-output process consists of two states and two symbol input and output alphabets, and its behaviour can be summarised as follows: the input-output process does nothing to the input whenever it sees a 0, while if it sees a 1, with probability $\alpha_j$ it will output a 0 if it is in the $j$-th state, and with probability $1-\alpha_j$ it will output a 1. If a 0 is output, a transition to state 1 happens, while if a 1 is output a transition to state 2.

The transition matrices, $T^{(y|x)}$, associated with the \etr{} are:
\begin{equation}
    \begin{aligned}
	T^{(0|0)} &= \begin{pmatrix}
		1 & 0 \\
	    1 & 0 
	\end{pmatrix} \,,
	&T^{(1|0)} &= \begin{pmatrix}
	0 & 0 \\
	0 & 0 
\end{pmatrix} \,, \notag \\
	T^{(0|1)} &= \begin{pmatrix}
	\alpha_1 & 0 \\
	1-\alpha_2 & 0 
\end{pmatrix} \,,
	&T^{(1|1)} &= \begin{pmatrix}
	0 & 1-\alpha_1 \\
	0 & \alpha_2 
\end{pmatrix} \,.
\end{aligned}
\end{equation}
The maximum fidelity constraint is equal to
\begin{align}
	F_{12} = \sqrt{\alpha_1 (1-\alpha_2)}+ \sqrt{\alpha_2 (1-\alpha_1)}.
\end{align}
A quantum model that saturates the fidelity constraint will minimise the quantum complexity and thus will be an optimal quantum model. Explicitly constructing the two quantum causal states as
\begin{align}
	&\ket{s_1} =  \sqrt{\alpha_1} \ket{0} +\sqrt{1-\alpha_1}\ket{1} \,, \notag \\
	&\ket{s_2} =    \sqrt{1-\alpha_2} \ket{0} +\sqrt{\alpha_2}\ket{1}\,,
\end{align}
it follows that $\braket{s_2}{s_1}=F_{12}$. Thus, this encoding corresponds to an optimal quantum model.

\begin{figure}[!t]
	\centering
    \scalebox{0.9}{
	\begin{tikzpicture}[node distance={27mm}, thick, main/.style = {draw, circle,minimum size=0.7cm}] 
		\node[main] (1) {$1$}; 
		\node[main] (2) [ right of=1] {$2$}; 
		
		\draw[->] (1) to [out=45,in=135,looseness=1] node[align=center,midway,above] {$0:1-(1-r)\alpha_2$ } (2); 
		
		\draw[->] (2) to [out=45,in=-45,looseness=5] node[align=center,midway,right] {$0:1-$ \\ $(1-r)(1-\alpha_1)$ } (2); 
		\draw[->] (2) to [out=-135,in=-45,looseness=1] node[align=center,midway,below] {$1:(1-r)(1-\alpha_1)$}  (1); 
		\draw[->] (1) to [out=225,in=135,looseness=5] node[align=center,midway,left] {$1:(1-r)\alpha_2$  } (1);
	\end{tikzpicture} }
	\caption{The Ising spin chain process.}
	\label{Fig:Ising}
\end{figure}
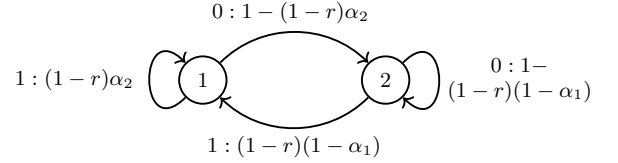

To obtain the resulting output process we first create the joint process over the joint alphabet of inputs and outputs and then trace over input symbols. Specifically, the output of this family of input-output processes for an IID input is found as
\begin{align}
	\tilde{F}^{(y)}=\sum_x T^{(y|x)} \otimes E^{(x)} = \sum_x T^{(y|x)} \cdot E^{(x)}\,,
\end{align}
with $E^{(x)}= \{r,1-r\}$ the probabilities of emitting a 0 or 1 from the IID process, and $T^{(y|x)}$ the transition matrices of the input-output process. We use the tilde symbol to signal that, in general, the output process may not be in its minimal presentation, i.e. its \ema{}. However, as this is not the case in the particular example we are considering, we drop the tilde to denote the \ema{}, $\tilde{F}^{(y)}=F^{(y)}$. The transition matrices of the output are explicitly
\begin{align}
	F^{(0)} &= \begin{pmatrix}
		0 & 1-\alpha_2(1-r) \\
		0 & 1-(1-\alpha_1)(1-r) 
	\end{pmatrix} \,, \, \,  \notag \\
	F^{(1)} &= \begin{pmatrix}
		\alpha_2(1-r) & 0 \\
		(1-\alpha_1)(1-r) & 0 
	\end{pmatrix} \,.
\end{align}
The behaviour of the output is essentially the Ising spin chain \cite{suen_classical-quantum_2017} shown in Fig.\@ \ref{Fig:Ising}, up to a redefinition of parameters.

The quantum causal states of the optimal quantum model for the Ising spin chain are
\begin{align}
	\ket{\sigma_1} &= \sqrt{1-(1-r)\alpha_2} \, \ket{0} + \sqrt{(1-r)\alpha_2} \,\ket{1} \,, \notag \\
	\ket{\sigma_2} &=   \sqrt{1-(1-r)(1-\alpha_1)} \, \ket{0} +\sqrt{(1-r)(1-\alpha_1)} \, \ket{1}.
\end{align}
It is straightforward to check that the overlap saturates the fidelity constraint, i.e. $\braket{\sigma_1}{\sigma_2}=F_{12}$, which shows that this is an optimal quantum model.

The stationary distributions of the \ema{} of the Ising spin chain and that of the \etr{} of the input-output process in Fig.\@ (\ref{Fig:Agent 3}), when driven by the biased coin, are the same. As a result, their classical complexities are also equal. This is not true, however, for their respective optimal quantum models, as the quantum causal state overlaps are different, and thus the quantum complexities differ.
\begin{figure}[!t]
	\centering
	\includegraphics[width=1\linewidth]{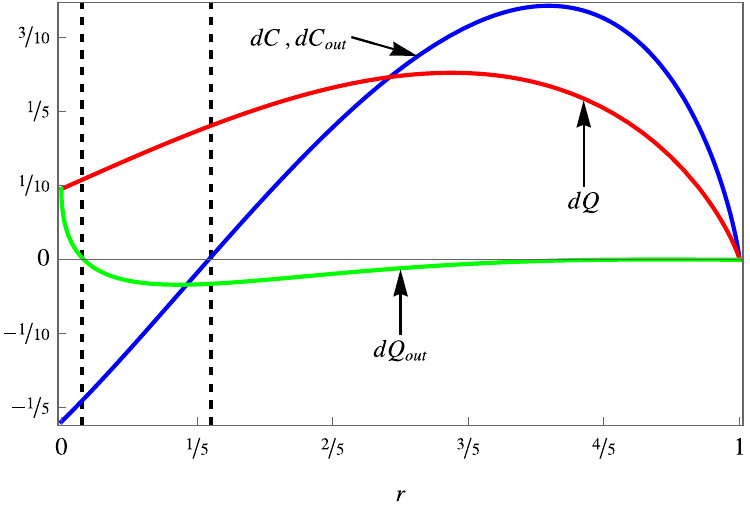}
	\caption{The difference of classical and quantum complexities of the two input-output processes and output processes as functions of the bias of the input coin $r$. In blue are both the difference of the classical complexities of the two input-output processes and the two outputs, $dC, dC_{out}$. In red we plot the difference of the quantum complexities of optimal quantum agents, $dQ$, while with green the difference of quantum complexities of optimal quantum models of the outputs, $dQ_{out}$. The vertical dashed lines separate the three regions of interest (see main text).  }
	\label{Fig:output vs transducer}
\end{figure}  

Consider two input-output processes from the family, one with parameter values $\alpha_1 =0, \alpha_2 =\nicefrac{7}{10}, $ which we denote by $\A_1$  and the other with $\alpha_1 = \nicefrac{2}{3}, \alpha_2 = \nicefrac{7}{10}$, denoted by $\A_2$.  We assume that they are reacting to inputs from a biased coin of bias $r$, denoting it with $\I$, while we denote its outputs with $\O_1$ and $\O_2$, respectively. We form the difference of the complexities of the optimal models of the agents and of the outputs, which we denote $dC=C^{\A_1}_\B - C^{\A_2}_\B $ and $dC_{out}=C^{\O_1} - C^{\O_2} $, respectively. As noted, these actually coincide for this input process. Similarly, we consider the difference of the quantum complexities of optimal models of the agents and outputs, and denote them with $dQ=Q^{\A_1}_\B - Q^{\A_2}_\B $ and $dQ_{out}=Q^{\O_1} - Q^{\O_2} $, respectively. In Fig.\@ \ref{Fig:output vs transducer} we plot these quantities as functions of the bias $r$ of the input process. With the dashed vertical lines we separate three regions of different behaviour that demonstrate the three of the four possible cases: (ii) in the leftmost region, the complexities of both the agents and outputs are inconsistent, (iii) in the rightmost region only the complexities of the outputs are inconsistent, and (iv) in the middle region, only the complexities of the agents are inconsistent.

\section{Discussion} 

The complexity of a strategy involving adaptive decision-making is often quantified by its statistical complexity - how much memory an agent needs to track about the past to realise designated decisions. Here we first introduce an informational quantity - the \emph{channel excess entropy} - that lower bounds this memory cost. Deploying this toolkit, we proved that the relative complexity of two decision-making tasks can reverse depending on whether agents executing them use quantum memory. We explicitly demonstrated this inconsistency in a broad range of scenarios, including cases where we compare the relative complexity of (i) executing two different strategies on the same input, (ii) executing the strategy operating on two different inputs,  and (iii) an agent that transforms one process to another versus its operational inverse. In each scenario, the strategy an agent would find easier to execute, quantified by the required memory costs, differed depending on its capability to process quantum information. 
Thus we conclude that what strategic tasks are more complex can change depending on whether we use classical or quantum agents.

Our results here generalise the recent finding regarding \emph{ambiguity of simplicity} in predictive modelling of stochastic processes~\cite{aghamohammadiAmbiguitySimplicityQuantum2017,ghafari_errortolerant_2022a}; where quantum and classical modelling result in different conclusions about which processes require the most past information to model. Indeed, our results here encompass such discoveries as the special case where the input is a trivial sequence of zeros. However, our work shows that the ambiguity of complexity for input-output strategies can exhibit much richer behaviour. Sec.\@ \ref{sec: outputs vs strategies}, for example, demonstrated that quantum and classical agents can still disagree on which task is more complex - even when the end output processes defined by the agents' output actions do not exhibit this ambiguity.

There are also a number of exciting future directions for research. In this work, we assumed that the goal is to execute a target strategy. In many realistic games, the goal rather, is to maximise some reward function. We anticipate that similar classical-quantum ambiguities would persist in such scenarios, and it would be interesting to identify specific examples. The second is in thermodynamics: Recent discoveries have demonstrated that the memory required of an agent has significant thermodynamic consequences -- the larger the memory, the higher the heat dissipation~\cite{thompson2025energetic}. As such, any difference in complexities will likely lead to different conclusions in the minimal energetic costs of transforming one process to another. Could certain transformations then be energetically costly in the classical regime, but spontaneous when quantum processes are allowed? Any such discoveries would certainly be exciting, likely helping us better understand how notions of reversibility and temporal asymmetry vary between classical and quantum observers.

\acknowledgments
The definition of the channel excess entropy, along with an alternative proof that it lower-bounds a channel's statistical complexity, and a visual proof of Proposition \ref{prop: decomposition of E for IOP} through information diagrams, were first introduced in the doctoral thesis \cite{barnett_mechanisms_2016} of one of the authors (NB). 

The authors thank Thomas Elliott and James Moran for comments on an earlier version of the manuscript. 

This work is supported by KIAS individual grant numbers CG086202 (S.K.) and CG085302 (H.K.) at the Korea Institute for Advanced Study. M.G. acknowledges support from the NRF Investigatorship Program (Award No. NRF-NRFI09-0010), the National Quantum Office, hosted in A*STAR, under its Centre for Quantum Technologies Funding Initiative (S24Q2d0009), grant FQXi-RFP-IPW-1903 (``Are quantum agents more energetically efficient at making predictions?") from the Foundational Questions Institute (FQxI) and Fetzer Franklin Fund (a donor-advised fund of Silicon Valley Community Foundation),  the Singapore Ministry of Education Tier 1 Grant RT4/23.

\appendix

\section{Proof of bound between excess entropy and classical complexity of a stochastic process}
\label{App: proof of EE and C for passive SP}

In this section, we give the proof of Proposition \ref{prop: E<=C SP}.
\begin{proof}
    The random variables $\olharp{Y}$ and $\orharp{Y}$ that describe the semi-infinite past and future sequences of outputs, as well as the random variable $S$ that describes the causal states, form a Markov chain
    \begin{align}
        \olharp{Y}\rightarrow S\rightarrow \orharp{Y} \,.
    \end{align}
    This is due to the fact that a given past $\olharp{y}=\ldots y_{-2}y_{-1}$ is uniquely mapped to a causal state $s_i$ through the causal state function $\epsilon$ according to $\epsilon(\olharp{y})=s_i \,, s_i \in \S$. Similarly, the next output $y_{t+1}$ is generated depending only on the current state $s_t$; by iterating, the whole future is produced.
    Thus $\olharp{Y}, \orharp{Y}$ and $S$ form a Markov chain, and the data processing inequality \cite{cover_elements_2005} gives: 
    \begin{align}
        I[\olharp{Y}; \orharp{Y}] \leq I[S; \orharp{Y}].
    \end{align}
    The left-hand side is the excess entropy $\ee$. Using the definition of the mutual information on the right-hand side, we have
    \begin{align}
        \ee \leq H[S] - H[S|\orharp{Y}]\,, 
    \end{align}
    where $H[\cdot]$ denotes the Shannon entropy. 
    Finally, from the positivity of conditional entropy and the definition of the statistical complexity, we have
    \begin{align}
        \ee \leq C \,.
    \end{align}
\end{proof}

\section{Proof of bound between excess entropy and quantum complexity of a stochastic process}
\label{App: proof of EE and Q for passive SP}
In this section, we give the proof of Proposition \ref{prop: E<=Q SP}.
\begin{proof}
    A quantum model consists of an encoding function $\epsilon_q$ that maps classical states of the \ema{} to quantum states, as well as an evolution that extracts the classical emissions. The $\epsilon$ function of the \ema{} maps infinite pasts, $\olharp{y}$, to the causal states $s_i\in \S$ of the \ema{}, which are subsequently mapped to the quantum states $\ketbra{s_i}{s_i} \in \S_q$ of the quantum model through the encoding function $\epsilon_q$. Finally, a quantum channel $\Lambda: \S \rightarrow \Y \times \S $ extracts the classical output and updates the memory state. Diagrammatically:
   \begin{align}
       \olharp{y}=\ldots &y_{-2} y_{-1} \,\, \overset{\epsilon}{\longrightarrow} \,\,  \epsilon(\olharp{y})= s_i \notag \\
       &\,\, \overset{\epsilon_q}{\longrightarrow} \,\, \epsilon_q(s_i) = \ketbra{s_i}{s_i} \overset{\Lambda}{\longrightarrow} y_0 \,.
   \end{align} 
    By iterating, the entire semi-infinite future can be generated. It follows from Holevo's bound \cite{holevo_bounds_1973,nielsen_quantum_2010} that the mutual information between the past and the future is upper-bounded by the Holevo quantity $\chi=S[\rho]-\sum_i \Pr(s_i) S[\ketbra{s_i}{s_i}]=S[\rho]$, where $\rho=\sum_i \Pr(s_i)\ketbra{s_i}{s_i}$ denotes the average state of the quantum memory.
    We thus have:
    \begin{align}
        I[\olharp{Y}; \orharp{Y}] \leq S[\rho].
    \end{align}
    The left-hand side is the excess entropy $\ee$, while the right-hand side is the quantum complexity $Q$, which gives the desired result, $\ee \leq Q$.
\end{proof}

\section{Proof of bound between excess entropy and classical complexity of an input-output process}
\label{App: proof of EE and C for passive IP}
In this section, we give the proof of Proposition \ref{prop: E<=C IOP}.
\begin{proof}
    The proof follows in spirit the one in the case of stochastic processes, but care needs to be taken as we are dealing with conditional quantities. 
    First we show that a conditional version of the data-processing inequality holds. Specifically, we need to prove that if three random variables $X,Y,Z$ form a Markov chain when conditioned on another random variable $W$,
    \begin{align}
        X\rightarrow Y\rightarrow Z \,\, | \,\, W \,,
    \end{align}
    then we have  the following relation
   \begin{align}
        I[X; Z | W] \leq I[X; Y | W]\,.
    \end{align}
    We employ the chain rule for the conditional mutual information 
    \begin{align}
        I[X;Y,Z] &= I[X;Z]+I[X;Y|Z] \notag \\
        &= I[X;Y]+I[X;Z|Y]\,,
    \end{align}
    and apply it twice in the mutual information $I[X;Y,Z,W]$ to obtain
    \begin{align}
        I[X;Y,Z|W] = I[X;Y|W]+I[X;Z|Y,W]\,, \label{eq: chain rule for CMI}
    \end{align}
    as well as
    \begin{align}
        I[X;Y,Z|W] = I[X;Z|W]+I[X;Y|Z,W]. \label{eq: chain rule for CMI, second version}
    \end{align}
    As by assumption we have a Markov chain, even after conditioning on $W$, the conditional probability factorises as $\Pr(X,Z|Y,W)=\Pr(X|Y,W)\Pr(Z|Y,W)$, which implies that $I[X;Z|Y,W]=0$. Substituting in Eq.\@ \ref{eq: chain rule for CMI}, and combining with Eq.\@ \ref{eq: chain rule for CMI, second version}, we obtain
    \begin{align}
        I[X;Z|W]+I[X;Y|Z,W] = I[X;Y|W] \,.
    \end{align}
    The non-negativity of conditional mutual information gives the desired result
    \begin{align}
         I[X;Y|W] \leq I[X;Z|W] \,. \label{eq: conditional DPI}
    \end{align}
    
    To prove the inequality for the excess entropy of the channel, we notice that the random variables $\olharp{(X,Y)}$ of the joint past of inputs and outputs, $\orharp{Y}$ of future sequences of outputs, and $S$ that describes the causal states, when conditioned on the random variable $\orharp{X}$ that describes the infinite future of inputs, form a Markov chain
    \begin{align}
        \olharp{(X,Y)}\rightarrow S \rightarrow \orharp{Y} \,\, | \,\, \orharp{X} \,.
    \end{align}
    This is due to the fact that a given joint past $\olharp{(x,y)}=\ldots (x_{-2},y_{-2}) (x_{-1},y_{-1})$ is uniquely mapped to a causal state $s_i$ through the causal state function $\epsilon$ according to $\epsilon\left(\olharp{(x,y)}\right)=s_i \,, s_i \in \S$. Conditioning on future inputs does not affect the causal state assignment. Then, conditional on the next input $x_1$, the next future output $y_1$ is generated. This leads to an updated past history, mapped to another causal state through the causal state function, and upon receiving the next input it is mapped to the next output. The whole future outputs can be generated this way. This shows that $\olharp{(X,Y)}$, $S$, and $\orharp{Y}$ form a Markov chained, conditioned on the entire future $\orharp{X}$.
    Thus, the conditional version of the data processing inequality Eq.\@ \eqref{eq: conditional DPI} gives:
    \begin{align}
        I[&\olharp{(X,Y)}; \orharp{Y}|\orharp{X}] \leq I[S; \orharp{Y}|\orharp{X}] \notag \\
        &= H[\olharp{(X,Y)}|\orharp{X}]-H[\olharp{(X,Y)}|\orharp{(X,Y)}].
    \end{align}
    The left-hand side is the excess entropy $\ee^\A_\I$ of the channel. From the positivity of conditional entropy we thus have 
    \begin{align}
        \ee^\A_\I \leq H[\olharp{(X,Y)}|\orharp{X}] = H[S|\orharp{X}] \leq H[S]\,, 
    \end{align}
    where we used the causal state function.
    When driven by a specific input process, the causal states are driven into a specific distribution $\pi_\I$. Thus, $H[S] = H[\pi_\I] = C^{\A}_\I$, and we finally obtain
    \begin{align}
        \ee^{\A}_\I \leq C^{\A}_\I \,.
    \end{align}
    
\end{proof}

\section{Proof of bound between excess entropy and quantum complexity of an input-output process}
\label{App: proof of EE and Q for passive IP}
In this section, we give the proof of Proposition \ref{prop: E<=Q IOP}.
\begin{proof}
    A quantum model consists of an encoding function $\epsilon_q$ that maps classical states of the \etr{} to quantum states, as well as an evolution that extracts the classical outputs, conditional on inputs. The $\epsilon$ function of the \ema{} maps infinite joint pasts of inputs and outputs, $\olharp{(x,y)}$, to the causal states $s_i\in \S$ of the \etr{}, which are subsequently mapped to the quantum states $\ketbra{s_i}{s_i} \in \S_q$ of the quantum model. Finally, a quantum channel $\Lambda: \X \times \S \rightarrow \Y \times \S $ extracts the classical output, given input, and updates the memory state. We highlight that the next output, depends only on the current state of the memory and the received input. Diagrammatically:
   \begin{align}
       \olharp{(x,y)}&=\ldots (x_{-2},y_{-2}) (x_{-1},y_{-1}) \notag \\
       &\,\, \overset{\epsilon}{\longrightarrow} \,\,  \epsilon(\olharp{(x,y)})= s_i \,\, \overset{\epsilon_q}{\longrightarrow} \,\, \epsilon_q(s_i) = \ketbra{s_i}{s_i}   \notag \\
       &\,\,\,\,\,\overset{x_o}{\longrightarrow}\ketbra{x_0}{x_0}\otimes\ketbra{s_i}{s_i} \overset{\Lambda}{\longrightarrow} y_0 \,.
   \end{align} 
    The difference from the case of passive stochastic processes is that for each semi-infinite future $\orharp{x}$, there is a particular stationary distribution $\{\Pr(s_i|\orharp{x})\}$ for the causal states $\{s_i\}$. It follows from Holevo's bound \cite{holevo_bounds_1973,nielsen_quantum_2010} that the mutual information between the joint past and the future outputs, given future inputs is upper-bounded by the Holevo quantity $\chi=S[\rho_{\,\orharp{x}}]-\sum_i \Pr(s_i|\orharp{x}) S[\ketbra{s_i}{s_i}]=S[\rho_{\,\orharp{x}}]$, where $\rho_{\,\orharp{x}}=\sum_i \Pr(s_i|\orharp{x})\ketbra{s_i}{s_i}$ denotes the average state of the quantum memory, given the semi-infinite future $\orharp{x}$.
    We thus have:
    \begin{align}
        I[\olharp{(X,Y)}; \orharp{Y} | \orharp{x}] \leq S[\rho_{\,\orharp{x}}]\,.
    \end{align}
    By averaging the last inequality over all possible futures, we have
    \begin{align}
        \sum_{\orharp{x}}\Pr(\orharp{x}) I[\olharp{(X,Y)}; \orharp{Y} | \orharp{x}]  \leq \sum_{\orharp{x}}\Pr(\orharp{x}) S[\rho_{\,\orharp{x}}]\,.
    \end{align}
    The left hand side, $I[\olharp{(X,Y)}; \orharp{Y} | \orharp{X}]$, is the definition of the channel excess entropy $\ee^\A_\I$, and thus we have
    \begin{align}
        \ee^\A_\I \leq \sum_{\orharp{x}}\Pr(\orharp{x}) S[\rho_{\,\orharp{x}}] \leq S\Big[\sum_{\orharp{x}}\Pr(\orharp{x}) \rho_{\,\orharp{x}}\Big]\,,
    \end{align}
    where the last inequality follows from the concavity of the von Neumann entropy. Using the definition of $\rho_{\orharp{x}}$ in the right hand side, we obtain
    \begin{align}
        &S\Big[\sum_{\orharp{x}}\Pr(\orharp{x}) \rho_{\,\orharp{x}}\Big]\notag \\
        &= S\Big[\sum_{\orharp{x}}\Pr(\orharp{x}) \sum_i \Pr(s_i|\orharp{x})\ketbra{s_i}{s_i}\Big] \notag \\
        &= S[\sum_i \Pr(s_i) \ketbra{s_i}{s_i}] = S[\rho]\,,
    \end{align}
    where we have used the definition of the average state of the quantum memory, $\rho=\sum_i \Pr(s_i) \ketbra{s_i}{s_i}$. Its von Neumann entropy gives the quantum complexity and thus we obtain the desired result
    \begin{align}
       \ee^\A_\I \leq Q^\A_\I \,.
    \end{align}
\end{proof}

\section{Proof of the decomposition of channel excess entropy}
\label{App: proof of decomposition of EE}
Here we give the proof of Proposition \ref{prop: decomposition of E for IOP}.
\begin{proof}
	We start by noting that the excess entropy of the process over joint inputs and outputs assumes the decomposition
	\begin{align}
		\ee^{\J} &\equiv I[\olharp{(X,Y)}; \orharp{(X,Y)}] 
		 = \ee^{\A}_\I +  \ee^{\I} + I[ \orharp{X};  \olharp{Y} |  \olharp{X}] \notag \\	
		&= I[\olharp{X}; \orharp{X}] + I[\olharp{(X,Y)}; \orharp{Y} |  \orharp{X} ] + I[ \orharp{X};  \olharp{Y} |  \olharp{X}]\,,
	\end{align}
	which can be shown by using the symmetry of the mutual information as well as the following identity 
	\begin{align}
		I[A;B|C] = I[A;B,C]- I[A;C] \,. 
	\end{align}
	Finally, it remains to show that the term $ I[ \orharp{X};  \olharp{Y} |  \olharp{X}]$ vanishes for causal channels. As this quantity is the mutual information between future inputs and past outputs, when conditioned on past inputs, it is zero for anticipation-free channels. Mathematically, this follows from the fact that causal channels' word probabilities must obey $\Pr (Y_{t:t+L}|\olrharp{X})=\Pr(Y_{t:t+L}|\olharp{X}_{t+L})$ \cite{barnett_computational_2015}. 
\end{proof}

\section{Two agents reacting to the same stimulus: derivation of complexities \label{app: 2 agents 1 stimulus}}

We provide details on the derivation of the classical and quantum complexities for the example in Sec.\@ \ref{sec: 2 agents 1 stimulus}.

As the input process is an IID process, i.e. a biased coin, it is straight forward to compute the stationary distribution over the causal states of the \etrs{}. We start from the delay channel (noiseless detector with delay) in Fig.\@ \ref{Fig: Alice Delay channel and Bob}. Let us denote with $\pi_1$ and $\pi_2$ the long-term probabilities of being at states 1 and 2, respectively. A simple way to find their values is by looking at the graph in Fig.\@ \ref{Fig: Alice Delay channel and Bob}, and noting the flow of probability through the arrows that go to each state. For instance, for state 1, there is the arrow with the transition `$0|0$', which starts and ends at state 1, as well as the arrow with the transition `$1|0$'. The probability of the former happening is equal to the probability of being in state 1, which is $\pi_1$ by definition, multiplied by the probability of receiving input `$0$', which is just $r$, as the process is IID. The latter, it is similarly equal to the probability of being in state 2, which is $\pi_2$, multiplied by the probability of receiving input `$0$', which is just $r$. Thus, the long-term probability $\pi_1$ of being at state 1 must be equal to the flow of probability to that state, from which we get
\begin{align}
    \pi_1&= \pi_1 r + \pi_2 (1-r)\,, \notag \\
    \pi_1&+\pi_2 =1 \,,
\end{align}
where the last condition is the fact that $\pi_1$ and $\pi_2$ form a probability distribution and thus sum to one. It follows that $(\pi_1,\pi_2)=(r,1-r)$ and thus the classical complexity if the binary entropy of the distribution
\begin{align}
    C_\I^\A = h(r)\,.
\end{align}
Repeating the process for Bob's strategy $\B$ shown in Fig.\@ \ref{Fig: Alice Delay channel and Bob} we similarly find
\begin{align}
    C_\I^\B = h(b)\,,
\end{align}
with $b=\nicefrac{1}{(1+(1-r)(1-\alpha))}$.

From the optimal classical description of the agents we can construct the optimal quantum models \cite{thompson_using_2017}, \cite{elliott_quantum_2022}.  First, it is easy to show that an optimal quantum model of Alice will have the same complexity as the classical one. This follows from the fact that the fidelity between the conditional infinite futures given current state, also called \emph{future morphs} of the two causal states \cite{shalizi_computational_2001,elliott_quantum_2022}, is equal to zero forcing any quantum model to have orthogonal causal states as well. This will not be the case in general for Bob however. In both cases, to prove the optimality of the quantum model we show that it saturates a \emph{maximum fidelity constraint} \cite{elliott_quantum_2022}. Optimality then follows from the fact that for two states, the von Neumann entropy is a decreasing function of the overlap between the states.
Concretely, for Alice, we have the classical-to-quantum state encoding will necessarily map the classical state into a pair of orthogonal quantum states, thus leading to the same complexity. This follows from a maximum fidelity constraint
\begin{align}
F_{12} = \min_R \sum_{\overrightarrow{y}}\sqrt{P(\overrightarrow{y}|\sigma_1 , \overrightarrow{x})P(\overrightarrow{y}|\sigma_2 , \overrightarrow{x})}\,,
\end{align}
where $R$ denotes any possible input strategy~\cite{elliott_quantum_2022}.
This constraint imposes an information-theoretic bound on the overlap of the quantum states of the model; if the constraint is violated, the quantum model can no longer reproduce the classical statistics correctly. In order to evaluate the maximum value of the fidelity, we use the results in Appendix D of Ref.\@ \cite{elliott_quantum_2022}. Specifically, we have that for each pair of states there is an optimal choice of input stimulus and solve an associated set of linear equations. That is,
\begin{align}
    &F_{ss^\prime} =  \notag \\ 
    &\min_x \sum_y \sqrt{T^{(y|x)}_{s, \lambda(x,y,s)}T^{(y|x)}_{s^\prime ,\lambda(x,y,s^\prime)}} F_{\lambda(x,y,s) \lambda(x,y,s^\prime)} \,,
\end{align}
where $\lambda(x,y,s)$ is the deterministic function that takes the input state $s$ and the input and output symbols, $x$ and $y$, and gives the next state. The deterministic nature of $\lambda$ follows from the unifilarity property of the \etr{}. By direct calculation, we find that for Alice $F_{12}=0$, which means that the quantum states have to be orthogonal.
Thus, a quantum model cannot lead to a memory reduction over the classical model and $Q_\I^\A=C_\I^\A$.

For Bob we find the quantum states through Eqs.\@ \eqref{eq:quantum encoding 1}-\eqref{eq:quantum encoding 2}, which can be encoded in a single qubit:
\begin{align}
	&\ket{\sigma_1} =  \sqrt{\alpha} \ket{0} +\sqrt{1-\alpha}\ket{1}  \notag \\
	&\ket{\sigma_2} =\ket{0} \,.
\end{align}
An explicit calculation of the maximum fidelity constraint gives 
\begin{align}
    F_{12}&= \min\big\{F_{11},\sqrt{\alpha}F_{11} \big\} \notag \\
    &= \min \{1,\sqrt{\alpha}\} = \sqrt{\alpha},
\end{align}
as, by definition, the fidelity of a state with itself is always 1, i.e. $F_{ss}=1$.
Thus, we find the value $F_{12}= \sqrt{\alpha}$. 
It is straightforward to check that the states of the quantum model obey $\bk{\sigma_1}{\sigma_2} =\sqrt{\alpha}$, thus saturating the fidelity constraint and minimising the von Neumann entropy: the quantum model is optimal.

Having obtained an optimal classical to quantum state encoding, we can now derive the quantum complexity.  In the case of Bob the quantum complexity differs from the classical one and is explicitly found to be 
\begin{align}
    Q_\I^\B = h\left(\frac{c-\sqrt{c+d}}{2c}\right)\,,
\end{align}
where $c=2+r-\alpha r$ and $d=-4r(2-4\alpha+2\alpha^2)$. Note that for Bob the classical and quantum complexities depend on both the bias of the coin $r$, as well as the parameter $\alpha$ of the family of input-output strategies.

\section{A single agent reacting to two stimuli: derivation of complexities \label{app: 1 agent 2 stimuli}}
We provide details on the derivation of the complexities in the main text.

For the classical complexities, it suffices to derive the stationary distribution of the \etr{} for the two inputs. As the two inputs are IID, the stationary distribution of the joint process is the same as that of the \etr{}, for each given input process. Specifically,
\begin{align}
    J_i^{(x,y)}= T^{(y|x)} \I_i^{(x)},
\end{align}
where $T^{(y|x)}$ represent the transition matrices of the \etr{} shown in Fig.\@ \ref{Fig:etr A to B} with the values $p_1=p_2=0, p_3=\nicefrac{4}{7}, q_2=\nicefrac{3}{5},q_3=\nicefrac{1}{100}$ and  $q_1$ a free parameter, and $\I_1^{(x)}=\{\nicefrac{2}{10},\nicefrac{1}{10},\nicefrac{7}{10}\}$ and $\I_2^{(x)}=\{\nicefrac{1}{10},\nicefrac{7}{10},\nicefrac{2}{10}\}$ denote the probabilities of inputs for each of the two input processes. We explicitly find
\begin{align}
    \pi_1 &= \frac{1}{296666+4035 q_1} \left\{ 30666,266000,4035 q_1 \right\} \,, \notag \\
    \pi_2 &= \frac{1}{230919+15715 q_1} \left\{ 13919,217000,15715 q_1\right\}\,,
\end{align}
where $\pi_i$ denotes the stationary distribution of the \etr{} when driven by input $\I_i$.
The classical complexities follow by taking the Shannon entropies of these distributions.

We now derive the optimal quantum models. We first evaluate the maximum fidelity constraint through
\begin{align}
    &F_{ss^\prime} =  \notag \\ 
    &\min_x \sum_y \sqrt{T^{(y|x)}_{s, \lambda(x,y,s)}T^{(y|x)}_{s^\prime ,\lambda(x,y,s^\prime)}} F_{\lambda(x,y,s) \lambda(x,y,s^\prime)}.
\end{align}
We first note that for states 1 and 3 we have $F_{fu}=0$, which also forces that $F_{eu}=0$. Finally,
\begin{align}
    F_{fe} =  \min \left\{\sqrt{\frac{4}{7}}, \sqrt{\frac{99(1-q_1)}{100}} , 1\right\} \notag \\
    \min \left\{\sqrt{\frac{4}{7}}, \sqrt{\frac{99(1-q_1)}{100}}\right\}
\end{align}
It remains to construct the quantum states that satisfy the above constraint and then show that they also minimise the von Neumann entropy.
From the fidelity constraints, the quantum causal states $\ket{\sigma_i}$ must obey $\braket{f}{u}=\braket{e}{u}=0$, while $\braket{f}{e}=c$ with $\abs{c}\in [0,F_{fe}]$. That is, we need to construct a triple of states obeying the above constraints with a value of $c$ that minimises the entropy.

Let $\pi=\{\pi_f,\pi_e,\pi_u\}$ denote the stationary distribution over the causal states of the \etr{} for a given input. Then, the quantum complexity can be calculated through the eigenvalues of the Gram matrix, which shares non-zero eigenvalues with the state $\rho = \sum_{x\in\{f,e,u\}} \pi_x \ketbra{x}{x}$ \cite{jozsa_distinguishability_2000}.  The Gram matrix in this case is 
\begin{align}
	G = \begin{pmatrix}
		\pi_f & \sqrt{\pi_f \pi_e} c & 0 \\
		 \sqrt{\pi_f \pi_e} c^* & \pi_e & 0 \\
		 0 & 0 & \pi_u  
	\end{pmatrix} \,,
\end{align}
and with $\pi_u=1-\pi_f+\pi_e$ its eigenvalues are found to be 
\begin{align}
	e_i =\{&1-\pi_f-\pi_e \,,  \notag \\
    &\frac{1}{2} \left( \pi_f+\pi_e \pm \sqrt{(\pi_1-\pi_e)^2+4\abs{c}^2 \pi_f \pi_e} \right)\}\,.
\end{align} 
For given values of the $\pi_x$, it can be shown that the von Neumann entropy is monotonically decreasing with $F_{fe}$. Specifically, the derivative of the von Neumann entropy $S$ with respect to $\abs{c}$ is
\begin{align}
    \frac{d S}{d \abs{c}} = \frac{-2 \abs{c} \pi_f \pi_e }{k}  \log\Big( \frac{\pi_f +\pi_e +k}{\pi_f +\pi_e - k} \Big)\,, 
\end{align}
where $k=\sqrt{(\pi_f -\pi_e)^2 +4 \abs{c}^2 \pi_f \pi_e}$. It follows that $\frac{d S}{d \abs{c}}$ and thus the minimum value of the von Neuman entropy is achieved for the maximum value of $\abs{c}$.
Thus the quantum complexity is minimised for the maximum allowed overlap, $\abs{c}=F_{fe}$. 
Next, we need to construct a set of quantum states that achieve these overlaps. It is straightforward to check that such a triple is
\begin{align}
	\ket{f}&=  \ket{1}, \notag \\
	\ket{e}&= F_{fe}\ket{1}+\sqrt{1-F_{fe}^2}\ket{2} \notag \\
	\ket{u}&= \ket{3},
\end{align}
where $\{\ket{j}\}_{j=1,\ldots,3}$ denotes an orthonormal basis.

\section{Input-output processes with equal classical and quantum complexities with values that range in all of $\mathbb{R}$ \label{app: input-output processes that fill gap}} 

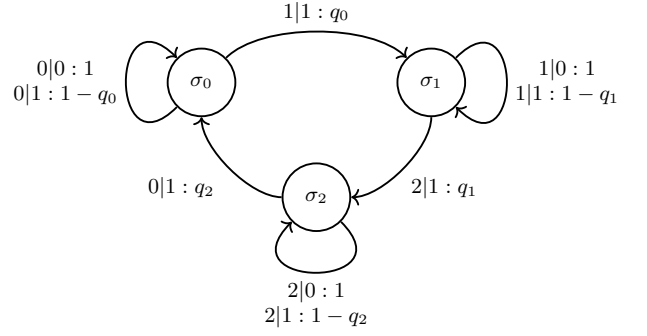
\begin{figure}[!ht]
	\centering
	\scalebox{0.9}{
		\begin{tikzpicture}[node distance={24mm}, thick, main/.style = {draw, circle,minimum size=1cm}] 
			\node[main] (1) {$\sigma_0$}; 
			\node[main] (3) [below right of=1] {$\sigma_2$}; 
			\node[main] (2) [above right of=3] {$\sigma_1$}; 
			
			\draw[->] (1) to [out=45,in=135,looseness=0.7] node[align=center, midway, above] {$1|1:q_{0}$}  (2); 
			\draw[->] (3) to [out=180,in=-90,looseness=0.7] node[align=center, midway, below left] {$0|1:q_2$ }  (1); 
			
			\draw[->] (2) to [in=0,out=-90,looseness=0.7] node[align=center, below right] {$2|1:q_1$ }  (3); 
			
			\draw[->] (1) to [out=225,in=135,looseness=5] node[align=center,midway,left] {$0|0:1$ \\ $0|1:1-q_0$} (1);
			\draw[->] (2) to [out=45,in=-45,looseness=5] node[align=center,midway,right] { $1|0:1$ \\ $1|1:1-q_1$} (2);
			\draw[->] (3) to [out=-45,in=-135,looseness=5] node[align=center,midway,below] {$2|0:1$ \\ $2|1:1-q_2$} (3);
			
		\end{tikzpicture} 
	}
	\caption{The \etr{} $\T_3$  from the family $\T_n$.}
	\label{Fig:etransducer covering R}
\end{figure}

Here we show that there exists a multiparameter family of input-output processes with equal classical and quantum complexities, with values that cover the whole of $\mathbb{R}$ when driven by an IID process. We explicitly construct such a family for the case of a two-symbol input alphabet $\X = \{0,1\}$, and an $n$-symbol output alphabet $\Y = \{0,\ldots, n-1\}$. In our considerations, the input process is taken as a biased coin, $\B$ with transition matrices $B^{(x)}=\{p,1-p\}$.

Let $\T_n$ denote the family of input-output processes, with $n\in \mathbb{N}$ denoting the number of causal states. The family of input-output processes has the following simple structure. At each state, input 0 leads to a self-transition back to the same state while outputting the label of the state. On the other hand, if in state $j$, input 1 leads to a self-transition while emitting the label of the state with probability $1-q_j$ or a transition to state $j+1$ (mod $n$) while emitting $j+1$ (mod $n$) with probability $q_j$. We show the input-output process with $n=3$ in Fig. \ref{Fig:etransducer covering R}.

For this family of input-output processes, the stationary distribution can be directly calculated and found to have the form
\begin{align}
    \pi_i = \frac{\prod_{j\neq i} q_j}{\sum_k\prod_{j\neq k} q_j} \,, \, \, \,  \forall i=0,\ldots,n-1\,,
\end{align}
where $\pi_i$ denotes the long term probability of being at state $i$. Due to the form of the $\pi_i$'s and the fact that all the $q_i$ can be freely chosen, it is clear that we can create any probability distribution of $n$ outcomes, $(\pi_0,\ldots, \pi_{n-1})$. Since the classical complexity is given by $C_\mu = -\sum_i \pi_i \log\pi_i$, we can thus reproduce any value of classical complexity in the range $(0,\log n)$. Thus, if there is an input-output process $\A$ with a gap between its classical and quantum complexities, $C^{(\A)}_q<C^{(\A)}_\mu$, we can choose an input-output process from the family $\T_n$ with $n$ such that $\log n \geq C^{(\A)}_\mu$ and then, by appropriately varying the parameters $q_i$, produce any value of complexity in the range $\left(C^{(\A)}_q,C^{(\A)}_\mu\right)$. It remains to show that the quantum complexities are the same as the classical complexities, which would then suffice to show that an ambiguity between classical and quantum complexities of the family of input-output processes and $\A$ exist for any value in the gap. The fact that no improvement in complexity can occur by any quantum model follows by noting that the maximum fidelity constraint \cite{elliott_quantum_2022} between any pair of quantum causal states of a quantum model is necessarily 0, which implies that the quantum states have to be orthogonal. With orthogonal states, however, classical and quantum complexities coincide. Specifically, the maximum fidelity constraint for states $i$ and $j$ are
\begin{align}
    F_{i,j} = \min\left\{ 0 \, , \, \,  \sqrt{q_i(1-q_j)} \delta_{j, i+1}  \right\} = 0 \,,
\end{align}
Thus, all quantum states have to be orthogonal which concludes the proof.

\section{Existence of input-output processes that can not exhibit ambiguous ordering for IID inputs \label{app: input output processes with no ambiguity IID}}
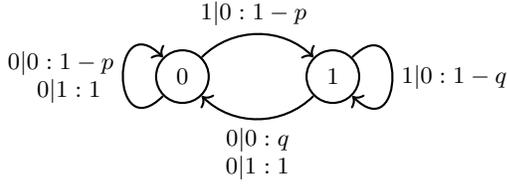
\begin{figure}[!h]
	\centering
	\begin{tikzpicture}[node distance={20mm}, thick, main/.style = {draw, circle,minimum size=0.7cm}] 
	\node[main] (1) {$0$}; 
	\node[main] (2) [ right of=1] {$1$}; 
			
	\draw[->] (1) to [out=45,in=135,looseness=1] node[align=center,midway,above] {$1|0:1-p$ } (2); 
	\draw[->] (2) to [out=-135,in=-45,looseness=1] node[align=center,midway,below] {$0|0:q$\\$0|1:1$}  (1); 
	\draw[->] (1) to [out=225,in=135,looseness=5] node[align=center,midway,left] {$0|0:1-p$ \\$\,\,\,\, 0|1:1$} (1);	
    \draw[->] (2) to [out=45,in=-45,looseness=5] node[align=center,midway,right] {$1|0:1-q$  } (2);
	\end{tikzpicture} 
	\caption{An input-output process that does not exhibit ambiguity of complexity for any pair of IID inputs.}
	\label{Fig:Appendix B agents}
\end{figure}
Consider the input-output process depicted in the Fig.\@ \ref{Fig:Appendix B agents}, whose action can be summarised as follows: from each state an input 1 leads to an output 0, while an input 0 leads to an output 0 or 1 with some probabilities, in general different. In addition, an emission is followed by a transition to the state of the same label.

A quantum model is one with quantum causal states
\begin{align}
	&\ket{s_0} =  \left( \sqrt{1-p} \ket{0}\ket{0} +\sqrt{p}\ket{1}\ket{1} \right) \ket{0}\ket{0} \,, \notag \\
	&\ket{s_1} =   \left( \sqrt{q} \ket{0}\ket{0} +\sqrt{1-q}\ket{1}\ket{1}\right)\ket{0}\ket{0}\,,
\end{align}
that have overlap
\begin{align}
    \braket{s_0}{s_1}=\sqrt{(1-p)q}+\sqrt{(p(1-q)}.
\end{align}
It is straight forward to check that the maximum fidelity constraint is equal to $F_{12}=\sqrt{(1-p)q}+\sqrt{(p(1-q)}=\braket{s_0}{s_1}$, which shows that the quantum model is optimal.

Consider now the case where the agent is driven by two biased coin processes $\B_r$ and $\B_{r'}$ with different biases $r$ and $r'$. For a biased coin process $\B_r$ the stationary distribution is readily obtained,
\begin{align}
    \varphi_0=\frac{1-(1-q)r}{1+(p+q-1)r} \, \,, \, \, \, \varphi_1 = 1-\varphi_0 \,.
\end{align}
Given that for a two-outcome probability distribution $\{\varphi_0, \varphi_1\}$, its entropy is lower the farther $\varphi_0$ is from the value $\nicefrac{1}{2}$ and assuming that $\varphi_1>\varphi_0>\min (\varphi_0^\prime,\varphi_1^\prime)$ and  $\varphi_1^\prime>\varphi_0^\prime$, it is easy to show that 
\begin{align}
   \varphi_0>\varphi_0^\prime \Longrightarrow r<r^\prime.
\end{align} 
Then, with $r<r^\prime$ and $\varphi_0>\varphi_0^\prime$, we readily obtain that the classical complexities of the agent when driven by the biased coins with biases $r$ and $r^\prime$ are
\begin{align}
    C_r<C_{r^\prime}\,.
\end{align}

For the quantum complexities, we first obtain the eigenvalues of the average memory state $\rho=\varphi_0 \ketbra{s_0}{s_0}+\varphi_1 \ketbra{s_1}{s_1}$, which are found to be
\begin{align}
    \lambda_{\pm}=\frac{1}{2}\pm \frac{\sqrt{1-4\varphi_0(1-\varphi_0)(1-F_{12}^2)}}{2}.
\end{align}
It is straightforward to show that in the case with $\nicefrac{1}{2}>\varphi_0>\varphi_0^\prime$ we have that $\lambda_+<\lambda_+^\prime$, which implies that
\begin{align}
    Q_r < Q_{r^\prime}\,,
\end{align}
showing that the quantum complexities are obeying the same ordering as the classical complexities. 

It remains to show this for the case with $\varphi_1>\varphi_0>\min (\varphi_0^\prime,\varphi_1^\prime)$ and $\min (\varphi_0^\prime,\varphi_1^\prime)=\varphi_1 ^\prime$. This follows, however, directly from a symmetry argument.

\section{Excess entropies for scenario A \label{app: excess entropies scenario A}}
We derive the excess entopies for scenario A in Section \ref{sec: 2 agents 1 stimulus}. 

As the excess entropy of the input is zero, i.e. $\ee_I = 0$, we have from Proposition \ref{prop: decomposition of E for IOP} that $\ee^\A_\I= \ee^\J-\ee^\I = \ee^\J$. As the input-output process $\A$ is deterministic, the excess entropy equals the classical (and quantum) complexity for any input. That is, 
\begin{align}
    \ee^{A}_{\I}&=C^{A}_{\I}=Q^{A}_{\I}=h(r)
     &= -\frac{(1-r)\log(1-r)+r \log r}{\log2} \,.
\end{align}

For the excess entropy $\ee^\B_\I$, we find

\begin{align}
    \ee^\B_\I =& \frac{(r-1)}{ (\alpha (r-1)-r+2)\log (2)} \cdot  \Bigg(  \log\left(\frac{1}{2-r+\alpha(r-1)}\right) \notag \\
   &+(\alpha-1)r \log\left(\frac{(1-\alpha)(1-r)}{2-r+\alpha(r-1)}\right) \notag \\
   &-\alpha \log\left(\frac{\alpha}{1+r(\alpha-1)}\right) \notag \\
   &-(\alpha-1) \log\left(\frac{1-r+\alpha(r-1)}{2-r+\alpha(r-1)}\right) \Bigg)\,.
\end{align}

\bibliography{AgentsAmbiguityBib.bib}

\begin{thebibliography}{32}%
\makeatletter
\providecommand \@ifxundefined [1]{%
 \@ifx{#1\undefined}
}%
\providecommand \@ifnum [1]{%
 \ifnum #1\expandafter \@firstoftwo
 \else \expandafter \@secondoftwo
 \fi
}%
\providecommand \@ifx [1]{%
 \ifx #1\expandafter \@firstoftwo
 \else \expandafter \@secondoftwo
 \fi
}%
\providecommand \natexlab [1]{#1}%
\providecommand \enquote  [1]{``#1''}%
\providecommand \bibnamefont  [1]{#1}%
\providecommand \bibfnamefont [1]{#1}%
\providecommand \citenamefont [1]{#1}%
\providecommand \href@noop [0]{\@secondoftwo}%
\providecommand \href [0]{\begingroup \@sanitize@url \@href}%
\providecommand \@href[1]{\@@startlink{#1}\@@href}%
\providecommand \@@href[1]{\endgroup#1\@@endlink}%
\providecommand \@sanitize@url [0]{\catcode `\\12\catcode `\$12\catcode
  `\&12\catcode `\#12\catcode `\^12\catcode `\_12\catcode `\%12\relax}%
\providecommand \@@startlink[1]{}%
\providecommand \@@endlink[0]{}%
\providecommand \url  [0]{\begingroup\@sanitize@url \@url }%
\providecommand \@url [1]{\endgroup\@href {#1}{\urlprefix }}%
\providecommand \urlprefix  [0]{URL }%
\providecommand \Eprint [0]{\href }%
\providecommand \doibase [0]{https://doi.org/}%
\providecommand \selectlanguage [0]{\@gobble}%
\providecommand \bibinfo  [0]{\@secondoftwo}%
\providecommand \bibfield  [0]{\@secondoftwo}%
\providecommand \translation [1]{[#1]}%
\providecommand \BibitemOpen [0]{}%
\providecommand \bibitemStop [0]{}%
\providecommand \bibitemNoStop [0]{.\EOS\space}%
\providecommand \EOS [0]{\spacefactor3000\relax}%
\providecommand \BibitemShut  [1]{\csname bibitem#1\endcsname}%
\let\auto@bib@innerbib\@empty
\bibitem [{\citenamefont {Barnett}\ and\ \citenamefont
  {Crutchfield}(2015)}]{barnett_computational_2015}%
  \BibitemOpen
  \bibfield  {author} {\bibinfo {author} {\bibfnamefont {N.}~\bibnamefont
  {Barnett}}\ and\ \bibinfo {author} {\bibfnamefont {J.~P.}\ \bibnamefont
  {Crutchfield}},\ }\bibfield  {title} {\bibinfo {title} {Computational
  {Mechanics} of {Input}–{Output} {Processes}: {Structured} {Transformations}
  and the {$\epsilon$}-{Transducer}},\ }\href
  {https://doi.org/10.1007/s10955-015-1327-5} {\bibfield  {journal} {\bibinfo
  {journal} {J Stat Phys}\ }\textbf {\bibinfo {volume} {161}},\ \bibinfo
  {pages} {404} (\bibinfo {year} {2015})}\BibitemShut {NoStop}%
\bibitem [{\citenamefont {Thompson}\ \emph {et~al.}(2017)\citenamefont
  {Thompson}, \citenamefont {Garner}, \citenamefont {Vedral},\ and\
  \citenamefont {Gu}}]{thompson_using_2017}%
  \BibitemOpen
  \bibfield  {author} {\bibinfo {author} {\bibfnamefont {J.}~\bibnamefont
  {Thompson}}, \bibinfo {author} {\bibfnamefont {A.~J.~P.}\ \bibnamefont
  {Garner}}, \bibinfo {author} {\bibfnamefont {V.}~\bibnamefont {Vedral}},\
  and\ \bibinfo {author} {\bibfnamefont {M.}~\bibnamefont {Gu}},\ }\bibfield
  {title} {\bibinfo {title} {Using quantum theory to simplify input–output
  processes},\ }\href {https://doi.org/10.1038/s41534-016-0001-3} {\bibfield
  {journal} {\bibinfo  {journal} {npj Quantum Inf}\ }\textbf {\bibinfo {volume}
  {3}},\ \bibinfo {pages} {1} (\bibinfo {year} {2017})},\ \bibinfo {note}
  {number: 1 Publisher: Nature Publishing Group}\BibitemShut {NoStop}%
\bibitem [{\citenamefont {Elliott}\ \emph {et~al.}(2022)\citenamefont
  {Elliott}, \citenamefont {Gu}, \citenamefont {Garner},\ and\ \citenamefont
  {Thompson}}]{elliott_quantum_2022}%
  \BibitemOpen
  \bibfield  {author} {\bibinfo {author} {\bibfnamefont {T.~J.}\ \bibnamefont
  {Elliott}}, \bibinfo {author} {\bibfnamefont {M.}~\bibnamefont {Gu}},
  \bibinfo {author} {\bibfnamefont {A.~J.}\ \bibnamefont {Garner}},\ and\
  \bibinfo {author} {\bibfnamefont {J.}~\bibnamefont {Thompson}},\ }\bibfield
  {title} {\bibinfo {title} {Quantum {Adaptive} {Agents} with {Efficient}
  {Long}-{Term} {Memories}},\ }\href
  {https://doi.org/10.1103/PhysRevX.12.011007} {\bibfield  {journal} {\bibinfo
  {journal} {Phys. Rev. X}\ }\textbf {\bibinfo {volume} {12}},\ \bibinfo
  {pages} {011007} (\bibinfo {year} {2022})},\ \bibinfo {note} {publisher:
  American Physical Society}\BibitemShut {NoStop}%
\bibitem [{\citenamefont {Aghamohammadi}\ \emph {et~al.}(2017)\citenamefont
  {Aghamohammadi}, \citenamefont {Mahoney},\ and\ \citenamefont
  {Crutchfield}}]{aghamohammadiAmbiguitySimplicityQuantum2017}%
  \BibitemOpen
  \bibfield  {author} {\bibinfo {author} {\bibfnamefont {C.}~\bibnamefont
  {Aghamohammadi}}, \bibinfo {author} {\bibfnamefont {J.~R.}\ \bibnamefont
  {Mahoney}},\ and\ \bibinfo {author} {\bibfnamefont {J.~P.}\ \bibnamefont
  {Crutchfield}},\ }\bibfield  {title} {\bibinfo {title} {The ambiguity of
  simplicity in quantum and classical simulation},\ }\href
  {https://doi.org/10.1016/j.physleta.2016.12.036} {\bibfield  {journal}
  {\bibinfo  {journal} {Physics Letters A}\ }\textbf {\bibinfo {volume}
  {381}},\ \bibinfo {pages} {1223} (\bibinfo {year} {2017})}\BibitemShut
  {NoStop}%
\bibitem [{\citenamefont {Ghafari}\ \emph {et~al.}(2022)\citenamefont
  {Ghafari}, \citenamefont {Gu}, \citenamefont {Ho}, \citenamefont {Thompson},
  \citenamefont {Suen}, \citenamefont {Wiseman},\ and\ \citenamefont
  {Pryde}}]{ghafari_errortolerant_2022a}%
  \BibitemOpen
  \bibfield  {author} {\bibinfo {author} {\bibfnamefont {F.}~\bibnamefont
  {Ghafari}}, \bibinfo {author} {\bibfnamefont {M.}~\bibnamefont {Gu}},
  \bibinfo {author} {\bibfnamefont {J.}~\bibnamefont {Ho}}, \bibinfo {author}
  {\bibfnamefont {J.}~\bibnamefont {Thompson}}, \bibinfo {author}
  {\bibfnamefont {W.~Y.}\ \bibnamefont {Suen}}, \bibinfo {author}
  {\bibfnamefont {H.~M.}\ \bibnamefont {Wiseman}},\ and\ \bibinfo {author}
  {\bibfnamefont {G.~J.}\ \bibnamefont {Pryde}},\ }\href
  {https://doi.org/10.48550/arXiv.1711.03661} {\bibinfo {title} {Error-tolerant
  witnessing of divergences in classical and quantum statistical complexity}}
  (\bibinfo {year} {2022}),\ \Eprint {https://arxiv.org/abs/1711.03661}
  {arXiv:1711.03661 [quant-ph]} \BibitemShut {NoStop}%
\bibitem [{\citenamefont
  {Crutchfield}(1994)}]{crutchfieldCalculiEmergenceComputation1994}%
  \BibitemOpen
  \bibfield  {author} {\bibinfo {author} {\bibfnamefont {J.~P.}\ \bibnamefont
  {Crutchfield}},\ }\bibfield  {title} {\bibinfo {title} {The calculi of
  emergence: Computation, dynamics and induction},\ }\href
  {https://doi.org/10.1016/0167-2789(94)90273-9} {\bibfield  {journal}
  {\bibinfo  {journal} {Physica D: Nonlinear Phenomena}\ }\textbf {\bibinfo
  {volume} {75}},\ \bibinfo {pages} {11} (\bibinfo {year} {1994})}\BibitemShut
  {NoStop}%
\bibitem [{\citenamefont {Crutchfield}\ and\ \citenamefont
  {Young}(1989)}]{crutchfield_inferring_1989}%
  \BibitemOpen
  \bibfield  {author} {\bibinfo {author} {\bibfnamefont {J.~P.}\ \bibnamefont
  {Crutchfield}}\ and\ \bibinfo {author} {\bibfnamefont {K.}~\bibnamefont
  {Young}},\ }\bibfield  {title} {\bibinfo {title} {Inferring statistical
  complexity},\ }\href {https://doi.org/10.1103/PhysRevLett.63.105} {\bibfield
  {journal} {\bibinfo  {journal} {Phys. Rev. Lett.}\ }\textbf {\bibinfo
  {volume} {63}},\ \bibinfo {pages} {105} (\bibinfo {year} {1989})},\ \bibinfo
  {note} {publisher: American Physical Society}\BibitemShut {NoStop}%
\bibitem [{\citenamefont {Shalizi}\ and\ \citenamefont
  {Crutchfield}(2001)}]{shalizi_computational_2001}%
  \BibitemOpen
  \bibfield  {author} {\bibinfo {author} {\bibfnamefont {C.~R.}\ \bibnamefont
  {Shalizi}}\ and\ \bibinfo {author} {\bibfnamefont {J.~P.}\ \bibnamefont
  {Crutchfield}},\ }\bibfield  {title} {\bibinfo {title} {Computational
  {Mechanics}: {Pattern} and {Prediction}, {Structure} and {Simplicity}},\
  }\href {https://doi.org/10.1023/A:1010388907793} {\bibfield  {journal}
  {\bibinfo  {journal} {Journal of Statistical Physics}\ }\textbf {\bibinfo
  {volume} {104}},\ \bibinfo {pages} {817} (\bibinfo {year}
  {2001})}\BibitemShut {NoStop}%
\bibitem [{\citenamefont {Crutchfield}(2012)}]{crutchfield_between_2012}%
  \BibitemOpen
  \bibfield  {author} {\bibinfo {author} {\bibfnamefont {J.~P.}\ \bibnamefont
  {Crutchfield}},\ }\bibfield  {title} {\bibinfo {title} {Between order and
  chaos},\ }\href {https://doi.org/10.1038/nphys2190} {\bibfield  {journal}
  {\bibinfo  {journal} {Nature Phys}\ }\textbf {\bibinfo {volume} {8}},\
  \bibinfo {pages} {17} (\bibinfo {year} {2012})},\ \bibinfo {note} {number: 1
  Publisher: Nature Publishing Group}\BibitemShut {NoStop}%
\bibitem [{Note1()}]{Note1}%
  \BibitemOpen
  \bibinfo {note} {Unifilarity is the property that guarantees determinism on
  the next state of the machine given the current state and
  emission}\BibitemShut {NoStop}%
\bibitem [{\citenamefont {Travers}\ and\ \citenamefont
  {Crutchfield}(2025)}]{travers_equivalence_2012}%
  \BibitemOpen
  \bibfield  {author} {\bibinfo {author} {\bibfnamefont {N.~F.}\ \bibnamefont
  {Travers}}\ and\ \bibinfo {author} {\bibfnamefont {J.~P.}\ \bibnamefont
  {Crutchfield}},\ }\bibfield  {title} {\bibinfo {title} {Equivalence of
  history and generator {$\epsilon$}-machines},\ }\bibfield  {journal}
  {\bibinfo  {journal} {Symmetry}\ }\textbf {\bibinfo {volume} {17}},\ \href
  {https://doi.org/10.3390/sym17010078} {10.3390/sym17010078} (\bibinfo {year}
  {2025})\BibitemShut {NoStop}%
\bibitem [{\citenamefont {Ellison}\ \emph {et~al.}(2009)\citenamefont
  {Ellison}, \citenamefont {Mahoney},\ and\ \citenamefont
  {Crutchfield}}]{ellison_prediction_2009}%
  \BibitemOpen
  \bibfield  {author} {\bibinfo {author} {\bibfnamefont {C.~J.}\ \bibnamefont
  {Ellison}}, \bibinfo {author} {\bibfnamefont {J.~R.}\ \bibnamefont
  {Mahoney}},\ and\ \bibinfo {author} {\bibfnamefont {J.~P.}\ \bibnamefont
  {Crutchfield}},\ }\bibfield  {title} {\bibinfo {title} {Prediction,
  {Retrodiction}, and the {Amount} of {Information} {Stored} in the
  {Present}},\ }\href {https://doi.org/10.1007/s10955-009-9808-z} {\bibfield
  {journal} {\bibinfo  {journal} {J Stat Phys}\ }\textbf {\bibinfo {volume}
  {136}},\ \bibinfo {pages} {1005} (\bibinfo {year} {2009})}\BibitemShut
  {NoStop}%
\bibitem [{\citenamefont {Gu}\ \emph {et~al.}(2012)\citenamefont {Gu},
  \citenamefont {Wiesner}, \citenamefont {Rieper},\ and\ \citenamefont
  {Vedral}}]{guQuantumMechanicsCan2012}%
  \BibitemOpen
  \bibfield  {author} {\bibinfo {author} {\bibfnamefont {M.}~\bibnamefont
  {Gu}}, \bibinfo {author} {\bibfnamefont {K.}~\bibnamefont {Wiesner}},
  \bibinfo {author} {\bibfnamefont {E.}~\bibnamefont {Rieper}},\ and\ \bibinfo
  {author} {\bibfnamefont {V.}~\bibnamefont {Vedral}},\ }\bibfield  {title}
  {\bibinfo {title} {Quantum mechanics can reduce the complexity of classical
  models},\ }\href {https://doi.org/10.1038/ncomms1761} {\bibfield  {journal}
  {\bibinfo  {journal} {Nat Commun}\ }\textbf {\bibinfo {volume} {3}},\
  \bibinfo {pages} {762} (\bibinfo {year} {2012})}\BibitemShut {NoStop}%
\bibitem [{Note2()}]{Note2}%
  \BibitemOpen
  \bibinfo {note} {Unifilarity of an $\epsilon $-transducer{} is the property
  that guarantees determinism on the next state of the machine given the
  current state, as well as current input and emission \cite
  {barnett_computational_2015}}\BibitemShut {NoStop}%
\bibitem [{\citenamefont {Suen}\ \emph {et~al.}(2017)\citenamefont {Suen},
  \citenamefont {Thompson}, \citenamefont {Garner}, \citenamefont {Vedral},\
  and\ \citenamefont {Gu}}]{suen_classical-quantum_2017}%
  \BibitemOpen
  \bibfield  {author} {\bibinfo {author} {\bibfnamefont {W.~Y.}\ \bibnamefont
  {Suen}}, \bibinfo {author} {\bibfnamefont {J.}~\bibnamefont {Thompson}},
  \bibinfo {author} {\bibfnamefont {A.~J.~P.}\ \bibnamefont {Garner}}, \bibinfo
  {author} {\bibfnamefont {V.}~\bibnamefont {Vedral}},\ and\ \bibinfo {author}
  {\bibfnamefont {M.}~\bibnamefont {Gu}},\ }\bibfield  {title} {\bibinfo
  {title} {The classical-quantum divergence of complexity in modelling spin
  chains},\ }\href {https://doi.org/10.22331/q-2017-08-11-25} {\bibfield
  {journal} {\bibinfo  {journal} {Quantum}\ }\textbf {\bibinfo {volume} {1}},\
  \bibinfo {pages} {25} (\bibinfo {year} {2017})}\BibitemShut {NoStop}%
\bibitem [{\citenamefont {Garner}\ \emph {et~al.}(2017)\citenamefont {Garner},
  \citenamefont {Liu}, \citenamefont {Thompson}, \citenamefont {Vedral},\ and\
  \citenamefont {Gu}}]{garner_provably_2017}%
  \BibitemOpen
  \bibfield  {author} {\bibinfo {author} {\bibfnamefont {A.~J.~P.}\
  \bibnamefont {Garner}}, \bibinfo {author} {\bibfnamefont {Q.}~\bibnamefont
  {Liu}}, \bibinfo {author} {\bibfnamefont {J.}~\bibnamefont {Thompson}},
  \bibinfo {author} {\bibfnamefont {V.}~\bibnamefont {Vedral}},\ and\ \bibinfo
  {author} {\bibfnamefont {m.}~\bibnamefont {Gu}},\ }\bibfield  {title}
  {\bibinfo {title} {Provably unbounded memory advantage in stochastic
  simulation using quantum mechanics},\ }\href
  {https://doi.org/10.1088/1367-2630/aa82df} {\bibfield  {journal} {\bibinfo
  {journal} {New Journal of Physics}\ }\textbf {\bibinfo {volume} {19}},\
  \bibinfo {pages} {103009} (\bibinfo {year} {2017})}\BibitemShut {NoStop}%
\bibitem [{\citenamefont {Ghafari}\ \emph {et~al.}(2019)\citenamefont
  {Ghafari}, \citenamefont {Tischler}, \citenamefont {Thompson}, \citenamefont
  {Gu}, \citenamefont {Shalm}, \citenamefont {Verma}, \citenamefont {Nam},
  \citenamefont {Patel}, \citenamefont {Wiseman},\ and\ \citenamefont
  {Pryde}}]{ghafari_dimensional_2019}%
  \BibitemOpen
  \bibfield  {author} {\bibinfo {author} {\bibfnamefont {F.}~\bibnamefont
  {Ghafari}}, \bibinfo {author} {\bibfnamefont {N.}~\bibnamefont {Tischler}},
  \bibinfo {author} {\bibfnamefont {J.}~\bibnamefont {Thompson}}, \bibinfo
  {author} {\bibfnamefont {M.}~\bibnamefont {Gu}}, \bibinfo {author}
  {\bibfnamefont {L.~K.}\ \bibnamefont {Shalm}}, \bibinfo {author}
  {\bibfnamefont {V.~B.}\ \bibnamefont {Verma}}, \bibinfo {author}
  {\bibfnamefont {S.~W.}\ \bibnamefont {Nam}}, \bibinfo {author} {\bibfnamefont
  {R.~B.}\ \bibnamefont {Patel}}, \bibinfo {author} {\bibfnamefont {H.~M.}\
  \bibnamefont {Wiseman}},\ and\ \bibinfo {author} {\bibfnamefont {G.~J.}\
  \bibnamefont {Pryde}},\ }\bibfield  {title} {\bibinfo {title} {Dimensional
  {{Quantum Memory Advantage}} in the {{Simulation}} of {{Stochastic
  Processes}}},\ }\href {https://doi.org/10.1103/PhysRevX.9.041013} {\bibfield
  {journal} {\bibinfo  {journal} {Phys. Rev. X}\ }\textbf {\bibinfo {volume}
  {9}},\ \bibinfo {pages} {041013} (\bibinfo {year} {2019})}\BibitemShut
  {NoStop}%
\bibitem [{\citenamefont {Elliott}\ \emph {et~al.}(2020)\citenamefont
  {Elliott}, \citenamefont {Yang}, \citenamefont {Binder}, \citenamefont
  {Garner}, \citenamefont {Thompson},\ and\ \citenamefont
  {Gu}}]{elliott_extreme_2020}%
  \BibitemOpen
  \bibfield  {author} {\bibinfo {author} {\bibfnamefont {T.~J.}\ \bibnamefont
  {Elliott}}, \bibinfo {author} {\bibfnamefont {C.}~\bibnamefont {Yang}},
  \bibinfo {author} {\bibfnamefont {F.~C.}\ \bibnamefont {Binder}}, \bibinfo
  {author} {\bibfnamefont {A.~J.~P.}\ \bibnamefont {Garner}}, \bibinfo {author}
  {\bibfnamefont {J.}~\bibnamefont {Thompson}},\ and\ \bibinfo {author}
  {\bibfnamefont {M.}~\bibnamefont {Gu}},\ }\bibfield  {title} {\bibinfo
  {title} {Extreme {{Dimensionality Reduction}} with {{Quantum Modeling}}},\
  }\href {https://doi.org/10.1103/PhysRevLett.125.260501} {\bibfield  {journal}
  {\bibinfo  {journal} {Phys. Rev. Lett.}\ }\textbf {\bibinfo {volume} {125}},\
  \bibinfo {pages} {260501} (\bibinfo {year} {2020})}\BibitemShut {NoStop}%
\bibitem [{\citenamefont {Thompson}\ \emph {et~al.}(2018)\citenamefont
  {Thompson}, \citenamefont {Garner}, \citenamefont {Mahoney}, \citenamefont
  {Crutchfield}, \citenamefont {Vedral},\ and\ \citenamefont
  {Gu}}]{thompson_causal_2018}%
  \BibitemOpen
  \bibfield  {author} {\bibinfo {author} {\bibfnamefont {J.}~\bibnamefont
  {Thompson}}, \bibinfo {author} {\bibfnamefont {A.~J.}\ \bibnamefont
  {Garner}}, \bibinfo {author} {\bibfnamefont {J.~R.}\ \bibnamefont {Mahoney}},
  \bibinfo {author} {\bibfnamefont {J.~P.}\ \bibnamefont {Crutchfield}},
  \bibinfo {author} {\bibfnamefont {V.}~\bibnamefont {Vedral}},\ and\ \bibinfo
  {author} {\bibfnamefont {M.}~\bibnamefont {Gu}},\ }\bibfield  {title}
  {\bibinfo {title} {Causal {Asymmetry} in a {Quantum} {World}},\ }\href
  {https://doi.org/10.1103/PhysRevX.8.031013} {\bibfield  {journal} {\bibinfo
  {journal} {Phys. Rev. X}\ }\textbf {\bibinfo {volume} {8}},\ \bibinfo {pages}
  {031013} (\bibinfo {year} {2018})},\ \bibinfo {note} {publisher: American
  Physical Society}\BibitemShut {NoStop}%
\bibitem [{\citenamefont {Kechrimparis}\ \emph {et~al.}(2023)\citenamefont
  {Kechrimparis}, \citenamefont {Gu},\ and\ \citenamefont
  {Kwon}}]{kechrimparis_causal_2023}%
  \BibitemOpen
  \bibfield  {author} {\bibinfo {author} {\bibfnamefont {S.}~\bibnamefont
  {Kechrimparis}}, \bibinfo {author} {\bibfnamefont {M.}~\bibnamefont {Gu}},\
  and\ \bibinfo {author} {\bibfnamefont {H.}~\bibnamefont {Kwon}},\ }\href
  {https://doi.org/10.48550/arXiv.2309.13572} {\bibinfo {title} {Causal
  {{Asymmetry}} of {{Classical}} and {{Quantum Autonomous Agents}}}} (\bibinfo
  {year} {2023}),\ \Eprint {https://arxiv.org/abs/2309.13572} {arXiv:2309.13572
  [quant-ph]} \BibitemShut {NoStop}%
\bibitem [{\citenamefont {Jozsa}\ and\ \citenamefont
  {Schlienz}(2000)}]{jozsa_distinguishability_2000}%
  \BibitemOpen
  \bibfield  {author} {\bibinfo {author} {\bibfnamefont {R.}~\bibnamefont
  {Jozsa}}\ and\ \bibinfo {author} {\bibfnamefont {J.}~\bibnamefont
  {Schlienz}},\ }\bibfield  {title} {\bibinfo {title} {Distinguishability of
  states and von {Neumann} entropy},\ }\href
  {https://doi.org/10.1103/PhysRevA.62.012301} {\bibfield  {journal} {\bibinfo
  {journal} {Phys. Rev. A}\ }\textbf {\bibinfo {volume} {62}},\ \bibinfo
  {pages} {012301} (\bibinfo {year} {2000})}\BibitemShut {NoStop}%
\bibitem [{\citenamefont {Holevo}(1973)}]{holevo_bounds_1973}%
  \BibitemOpen
  \bibfield  {author} {\bibinfo {author} {\bibfnamefont {A.~S.}\ \bibnamefont
  {Holevo}},\ }\bibfield  {title} {\bibinfo {title} {Bounds for the quantity of
  information transmitted by a quantum communication channel},\ }\href
  {http://mi.mathnet.ru/ppi903} {\bibfield  {journal} {\bibinfo  {journal}
  {Problemy Peredachi Informatsii}\ }\textbf {\bibinfo {volume} {9}},\ \bibinfo
  {pages} {3} (\bibinfo {year} {1973})},\ \bibinfo {note} {in Russian. English
  translation: Problems of Information Transmission, vol.~9, no.~3,
  pp.~177--183, 1973}\BibitemShut {NoStop}%
\bibitem [{\citenamefont {Nielsen}\ and\ \citenamefont
  {Chuang}(2010)}]{nielsen_quantum_2010}%
  \BibitemOpen
  \bibfield  {author} {\bibinfo {author} {\bibfnamefont {M.~A.}\ \bibnamefont
  {Nielsen}}\ and\ \bibinfo {author} {\bibfnamefont {I.~L.}\ \bibnamefont
  {Chuang}},\ }\href@noop {} {\emph {\bibinfo {title} {Quantum Computation and
  Quantum Information: 10th Anniversary Edition}}}\ (\bibinfo  {publisher}
  {Cambridge University Press},\ \bibinfo {year} {2010})\BibitemShut {NoStop}%
\bibitem [{\citenamefont {Cover}\ and\ \citenamefont
  {Thomas}(2005)}]{cover_elements_2005}%
  \BibitemOpen
  \bibfield  {author} {\bibinfo {author} {\bibfnamefont {T.~M.}\ \bibnamefont
  {Cover}}\ and\ \bibinfo {author} {\bibfnamefont {J.~A.}\ \bibnamefont
  {Thomas}},\ }\href {https://doi.org/10.1002/047174882X} {\emph {\bibinfo
  {title} {Elements of {Information} {Theory}}}}\ (\bibinfo  {publisher} {John
  Wiley \& Sons, Ltd},\ \bibinfo {year} {2005})\BibitemShut {NoStop}%
\bibitem [{\citenamefont {Crutchfield}\ \emph {et~al.}(2009)\citenamefont
  {Crutchfield}, \citenamefont {Ellison},\ and\ \citenamefont
  {Mahoney}}]{crutchfield_times_2009}%
  \BibitemOpen
  \bibfield  {author} {\bibinfo {author} {\bibfnamefont {J.~P.}\ \bibnamefont
  {Crutchfield}}, \bibinfo {author} {\bibfnamefont {C.~J.}\ \bibnamefont
  {Ellison}},\ and\ \bibinfo {author} {\bibfnamefont {J.~R.}\ \bibnamefont
  {Mahoney}},\ }\bibfield  {title} {\bibinfo {title} {Time's {Barbed} {Arrow}:
  {Irreversibility}, {Crypticity}, and {Stored} {Information}},\ }\href
  {https://doi.org/10.1103/PhysRevLett.103.094101} {\bibfield  {journal}
  {\bibinfo  {journal} {Phys. Rev. Lett.}\ }\textbf {\bibinfo {volume} {103}},\
  \bibinfo {pages} {094101} (\bibinfo {year} {2009})},\ \bibinfo {note}
  {publisher: American Physical Society}\BibitemShut {NoStop}%
\bibitem [{\citenamefont {Knoll}(2010)}]{knoll_radiation_2010}%
  \BibitemOpen
  \bibfield  {author} {\bibinfo {author} {\bibfnamefont {G.~F.}\ \bibnamefont
  {Knoll}},\ }\href@noop {} {\emph {\bibinfo {title} {Radiation {{Detection}}
  and {{Measurement}}}}}\ (\bibinfo  {publisher} {John Wiley \& Sons},\
  \bibinfo {year} {2010})\BibitemShut {NoStop}%
\bibitem [{\citenamefont {Migdall}(2013)}]{migdall_singlephoton_2013}%
  \BibitemOpen
  \bibfield  {author} {\bibinfo {author} {\bibfnamefont {A.}~\bibnamefont
  {Migdall}},\ }\href@noop {} {\emph {\bibinfo {title} {Single-Photon
  Generation and Detection}}},\ Experimental Methods in the Physical Sciences,
  Volume 45\ (\bibinfo  {publisher} {Academic Press},\ \bibinfo {address}
  {Waltham, MA},\ \bibinfo {year} {2013})\BibitemShut {NoStop}%
\bibitem [{\citenamefont {Ellison}\ \emph {et~al.}(2011)\citenamefont
  {Ellison}, \citenamefont {Mahoney}, \citenamefont {James}, \citenamefont
  {Crutchfield},\ and\ \citenamefont {Reichardt}}]{ellison_information_2011}%
  \BibitemOpen
  \bibfield  {author} {\bibinfo {author} {\bibfnamefont {C.~J.}\ \bibnamefont
  {Ellison}}, \bibinfo {author} {\bibfnamefont {J.~R.}\ \bibnamefont
  {Mahoney}}, \bibinfo {author} {\bibfnamefont {R.~G.}\ \bibnamefont {James}},
  \bibinfo {author} {\bibfnamefont {J.~P.}\ \bibnamefont {Crutchfield}},\ and\
  \bibinfo {author} {\bibfnamefont {J.}~\bibnamefont {Reichardt}},\ }\bibfield
  {title} {\bibinfo {title} {Information symmetries in irreversible
  processes},\ }\href {https://doi.org/10.1063/1.3637490} {\bibfield  {journal}
  {\bibinfo  {journal} {Chaos}\ }\textbf {\bibinfo {volume} {21}},\ \bibinfo
  {pages} {037107} (\bibinfo {year} {2011})}\BibitemShut {NoStop}%
\bibitem [{\citenamefont {James}\ \emph {et~al.}(2014)\citenamefont {James},
  \citenamefont {Mahoney}, \citenamefont {Ellison},\ and\ \citenamefont
  {Crutchfield}}]{james_many_2014}%
  \BibitemOpen
  \bibfield  {author} {\bibinfo {author} {\bibfnamefont {R.~G.}\ \bibnamefont
  {James}}, \bibinfo {author} {\bibfnamefont {J.~R.}\ \bibnamefont {Mahoney}},
  \bibinfo {author} {\bibfnamefont {C.~J.}\ \bibnamefont {Ellison}},\ and\
  \bibinfo {author} {\bibfnamefont {J.~P.}\ \bibnamefont {Crutchfield}},\
  }\bibfield  {title} {\bibinfo {title} {Many roads to synchrony: {Natural}
  time scales and their algorithms},\ }\href
  {https://doi.org/10.1103/PhysRevE.89.042135} {\bibfield  {journal} {\bibinfo
  {journal} {Phys. Rev. E}\ }\textbf {\bibinfo {volume} {89}},\ \bibinfo
  {pages} {042135} (\bibinfo {year} {2014})},\ \bibinfo {note} {publisher:
  American Physical Society}\BibitemShut {NoStop}%
\bibitem [{\citenamefont {Travers}\ and\ \citenamefont
  {Crutchfield}(2011)}]{travers_exact_2011}%
  \BibitemOpen
  \bibfield  {author} {\bibinfo {author} {\bibfnamefont {N.~F.}\ \bibnamefont
  {Travers}}\ and\ \bibinfo {author} {\bibfnamefont {J.~P.}\ \bibnamefont
  {Crutchfield}},\ }\bibfield  {title} {\bibinfo {title} {Exact
  {Synchronization} for {Finite}-{State} {Sources}},\ }\href
  {https://doi.org/10.1007/s10955-011-0342-4} {\bibfield  {journal} {\bibinfo
  {journal} {J Stat Phys}\ }\textbf {\bibinfo {volume} {145}},\ \bibinfo
  {pages} {1181} (\bibinfo {year} {2011})}\BibitemShut {NoStop}%
\bibitem [{\citenamefont {Thompson}\ \emph {et~al.}(2025)\citenamefont
  {Thompson}, \citenamefont {Riechers}, \citenamefont {Garner}, \citenamefont
  {Elliott},\ and\ \citenamefont {Gu}}]{thompson2025energetic}%
  \BibitemOpen
  \bibfield  {author} {\bibinfo {author} {\bibfnamefont {J.}~\bibnamefont
  {Thompson}}, \bibinfo {author} {\bibfnamefont {P.~M.}\ \bibnamefont
  {Riechers}}, \bibinfo {author} {\bibfnamefont {A.~J.}\ \bibnamefont
  {Garner}}, \bibinfo {author} {\bibfnamefont {T.~J.}\ \bibnamefont
  {Elliott}},\ and\ \bibinfo {author} {\bibfnamefont {M.}~\bibnamefont {Gu}},\
  }\bibfield  {title} {\bibinfo {title} {Energetic advantages for quantum
  agents in online execution of complex strategies},\ }\href@noop {} {\bibfield
   {journal} {\bibinfo  {journal} {arXiv preprint arXiv:2503.19896}\ }
  (\bibinfo {year} {2025})}\BibitemShut {NoStop}%
\bibitem [{\citenamefont {Barnett}(2016)}]{barnett_mechanisms_2016}%
  \BibitemOpen
  \bibfield  {author} {\bibinfo {author} {\bibfnamefont {N.}~\bibnamefont
  {Barnett}},\ }\href@noop {} {\emph {\bibinfo {title} {Mechanisms within the
  Black Box: Prediction, Computation, Randomness, and Complexity of
  Input-Output Processes via the $\varepsilon$-Transducer}}}\ (\bibinfo
  {publisher} {Unpublished doctoral dissertation, University of California},\
  \bibinfo {year} {2016})\BibitemShut {NoStop}%
\end{thebibliography}%

\end{document}